\documentclass[journal,twoside,web]{ieeecolor}
\usepackage{generic}
\usepackage{cite}
\usepackage{amsmath,amssymb,amsfonts}
\usepackage{algorithmic}
\usepackage{graphicx}
\usepackage{textcomp}
\usepackage{diagbox}

\newtheorem{Theorem}{Theorem}

\newtheorem{Lemma}{Lemma}
\newtheorem{Corollary}{Corollary}
\newtheorem{step1}{Step}
\newtheorem{step2}{Step}

\newtheorem{case}{Case}

\newtheorem{definition}{Definition}
\newtheorem{example}{Example}
\newtheorem{assumption}{Assumption}
\newtheorem{remark}{Remark}

\def\BibTeX{{\rm B\kern-.05em{\sc i\kern-.025em b}\kern-.08em
    T\kern-.1667em\lower.7ex\hbox{E}\kern-.125emX}}
\begin{document}
\title{Transmission-Constrained Consensus of Multiagent Networks}
\author{Xiaotian Wang, Housheng Su
\thanks{This work was supported by the National Natural Science Foundation of China under Grant Nos. 61991412, 61873318, and U1913602, and the Program for HUST Academic Frontier Youth Team under Grant No. 2018QYTD07.}
\thanks{The authors are with the School of Artificial Intelligence and Automation, Huazhong University of Science and Technology, and also with the Key Laboratory of Image Processing and Intelligent Control of Education Ministry of China, Wuhan 430074, China.
Email: XiaotianWangEmail@gmail.com, houshengsu@gmail.com.}}

\maketitle

\begin{abstract}
This paper studies the consensus problem for multiagent systems with transmission constraints. A novel model of multiagent systems is proposed where the information transmissions between agents are disturbed by irregular distortions or interferences (named transmission constraint functions), and this model is universal which can be applied in many cases, such as interval consensus and discarded consensus. In the transmission-constrained consensus problem, we obtain the necessary and sufficient condition that agents can converge to state consensus. Furthermore, a more general case is studied in which the system reaches an equilibrium. Based on some techniques of algebraic topology and stability theory, the existence, uniqueness and stability of the system equilibrium point can be proven, which means the system can reach an asymptotically stable equilibrium. Moreover, the state values of the  equilibrium are only decided by the network structure and transmission constraint functions, but not the agents' initial states. Finally, numerical simulations are presented to illustrate the proposed theorems and corollaries.
\end{abstract}


\section{Introduction}
\IEEEPARstart{I}{n} the past few years, distributed coordination of multiagent systems (MASs) has attracted much attention due to its broad application prospects in civil, military and other fields. One of its fundamental problems is consensus, which requires that agents achieve agreement about certain quantities of interest that depends on all agents’ states. Many scientific problems of consensus have emerged, and lots of control protocols are proposed in this area, such as consensus tracking \cite{hong2006tracking}, average consensus \cite{8361902} and robust consensus \cite{shi2013robust,8425632}.

In most of the above consensus problems, agents' states are not constrained. However, there are various state constraints on agents in many real-world scenarios, such as restricted actuators and limited communication distance. Imposing state constraints on agents has notable significance and great research value. The constrained consensus problems have been studied from different perspectives. For example, the constrained consensus and optimization problems have been studied, where agents' states are constrained in closed convex sets \cite{nedic2010constrained}. A novel state-constrained consensus (named interval consensus) problem has been proposed and studied in \cite{fontan2019interval}. Moreover, alternative approaches for imposing state constraints have been proposed in \cite{meng2016consensus,sun2013consensus}.

Information sharing is a necessary condition for MASs to admit a consensus solution. In the process of information transmitting, there are various unfavorable factors such as environmental interference, noise, and attenuation. Therefore, it is impractical and ideal to assume that agents can receive neighbors' information without distortions and noise disturbances. Previous researches on the imperfect information transmission often focused on switching topology, communication delay, communication link fault, packet loss, etc. For example, average consensus problem is studied in \cite{wu2012average}, where time-varying delay and packet loss are considered under the undirected communication network.

In the above constrained consensus problems, the constraints are imposed on agents' states or inputs, directly. In this work, we impose constraints on the information transmitting to make the agent's received information different from the original information of its neighbour, and research the effect of this difference on system stability.
To distinguish our problem from other constrained consensus problems, we call it \textbf{transmission-constrained consensus}.
In this problem, the deformed transmitted information is depicted by heterogeneous functions (named transmission constraint function). And a variety of functions can be chosen as transmission constraint function,
hence this study is so universal that it can be applied in many cases. Some applications and motivating examples are given in subsection \ref{subsc:applications}.

The first part of this work can be regarded as a study on the consensus conditions for the MAS with distorted transmitted state information (attenuation or saturation). The necessary and sufficient conditions for the ranges of transmission constraint functions are obtained.
A non-empty intersection of constraints is an important condition for MASs to achieve consensus. In many state-constrained consensus problems, it is assumed that the constraints have a non-empty intersection.
However, in this paper, we also consider the non-empty intersection case, where the multiagent systems may not achieve consensus, but an equilibrium.
In the second part of this study, it is proved that when the transmission constraint functions are distributed in a specific range (i.e., satisfy the conditions of Theorem \ref{th:converge_to_equilibrium} or \ref{th:single equilibrium}), the system will reach an asymptotically stable equilibrium even though that intersection of constraints is empty.

Compared with the existing works about the consensus of networked systems with constraints, the contributions of this work can be obtained as follows.
\begin{enumerate}
\item This work first studies the transmission-constrained consensus of multiagent networks. The transmission-constrained consensus model studied in this paper does not have a definite form, so that it can be regarded as a paradigm. The multiagent systems that can be translated to this model are able to make the system achieve consensus under the necessary conditions, such as interval consensus \cite{fontan2019interval}.
Unlike traditional constrained consensus problems, transmission-constrained consensus problem has the following features:
\begin{enumerate}
\item each link in the interaction network is limited by an individual constraint function, which is more general in reality;
\item constraint functions do not have a uniform type, and they can be various functions, such as trigonometric function, saturation function and Sigmoid function, etc.
\end{enumerate}
Those features make the transmission-constrained model have a wide range of application, but also bring heterogeneity into the dynamics, which increases the difficulty of analysis.
\item For the transmission-constrained consensus problem, we obtain some consensus conditions, in which a necessary and sufficient condition limits the distribution of constraint functions. As a more general case than the consensus case, equilibrium of MAS is seldom studied. We investigate this phenomenon and obtain conditions of equilibrium’s existence, uniqueness and stability.
Due to the novel model, where the unknown transmission constraints make the dynamics nonlinear, the analysis of MAS's stability is quite a challenge.
We design some linear boundaries and propose the corresponding lemmas to analyze the boundedness of dynamics. Then, we analyze the limit points of multiple solutions to prove the convergence of MASs. By coordinate transformations, another Lyapunov function is constructed to study the stability and uniqueness of equilibrium.
\end{enumerate}

The paper is organized as follows. Preliminaries and problem statement are given in Section II. Main results are provided in Section III. Supports of numerical examples are provided in Section IV, and Section V concludes this work. Finally, we put all proofs in appendixes.

\section{PRELIMINARIES}

\subsection{Graph Theory}
This paper studies the problem of transmission-constrained consensus of multiagent networks. Consider a MAS with $n$ agents, and denote $\mathbf{N}=\{1,2,\dots,n\}$. The finite vertex set is denoted by $\mathcal{V}=\{v_1,\dots,v_n\}$, and $\mathcal{E}\subseteq \mathcal{V}\times\mathcal{V}$ denotes edge set where $(v_j,v_i)\in \mathcal{E}$ means that there exists a communication link from agent $j$ to agent $i$. The adjacency weight matrix $\mathcal{A}\in\mathbb{R}^{n\times n}$ is defined as $a_{ij}>0$ if and only if $(v_j,v_i)\in \mathcal{E}$, and $a_{ij}=0$, otherwise. Then, the underlying interaction network of MAS is described by a (weighted) graph $\mathcal{G}=\{\mathcal{V},\mathcal{E},\mathcal{A}\}$ which is a triple. $(v_j,v_i)$ is defined as the directed edge from agent $j$ to agent $i$, and $\mathcal{N}_i=\{v_j\in \mathcal{V}:(v_j,v_i)\in\mathcal{E}\}$ denotes the neighbor set of agent $i$.
Denote $\alpha_i=\sum_{j=1}^na_{ij}$ as the row sum of $\mathcal{A}$, and $\bar{a}=\max \alpha_i$.

\subsection{Problem Statement}
For any $i\in\mathbf{N}$, denote the state of $v_i$ by $x_i(t)\in\mathbb{R}$. Then consider the continuous-time dynamics of single-integrator MAS with $n$ agents: $\dot{x}_i(t)=u_i(t)$, $i\in\mathbf{N}$, where $u_i(t)\in\mathbb{R}$ is the control input.

The problem studied in this work is different from the general MAS dynamics. In this problem, the information transmissions between agents are disturbed by interference functions (or attenuation functions), i.e., the transmission of state $x_i(t)$ is replaced by a transmission constrain function $f_{ij}\big(x_i(t)\big)$. $f_{ij}$ represents the transmission constraint imposed on the link from $i$ to $j$. Then, the transmission-constrained consensus algorithm of $x_i(t)$ is
\begin{equation}
\dot{x}_i(t)=u_i(t)=\sum_{j\in \mathcal{N}_i}a_{ij}\Big[f_{ji}\big(x_j(t)\big)-x_i(t)\Big].
\label{MAS_1}
\end{equation}

\begin{assumption}\label{assumption_1}
$\forall\,i,j\in\mathbf{N}$, the transmission constrain functions $f_{ij}(x)$ are piecewise continuous.
\end{assumption}

\begin{remark}
Since $\forall\,i,j\in\mathbf{N}$, $f_{ij}(x)$ are piecewise continuous, the autonomous system \eqref{MAS_1} may have multiple solutions. However, the following theorems and corollaries apply to both unique and multiple solutions.
\end{remark}

\begin{assumption}\label{assumption_2}
The directed graph $\mathcal{G}$ is strongly connected.
\end{assumption}

\begin{assumption}\label{assumption_3}
There exists an interval $[\partial_m,\partial_M]$, a value $\partial\in[\partial_m,\partial_M]$ and two rays
\begin{align*}
L_{1}(x)&=k_{1}(x-\partial)+\partial,\quad x\in(-\infty,\partial];\\
L_{2}(x)&=k_{2}(x-\partial)+\partial,\quad x\in[\partial,+\infty),
\end{align*}
where $k_{1},k_{2}<0$, such that $\forall\,j\in\mathbf{N},\,i\in\mathcal{N}_j$,
\begin{equation*}
\begin{alignedat}{2}
&x\le f_{ij}(x)< L_{1}(x),\quad & &x\in(-\infty,\partial_m);\\
&L_{2}(x)< f_{ij}(x)\le x, & &x\in(\partial_M,+\infty).
\end{alignedat}
\end{equation*}
\end{assumption}

\begin{assumption}\label{assumption_4}
For any $x'\in(-\infty,\partial_m)\cup(\partial_M,+\infty)$, there eixst $j\in \mathbf{N}$ and $i\in\mathcal{N}_j$, such that $f_{ij}(x)$ is continuous on $x'$ and $f_{ij}(x')\ne x'$.
\end{assumption}

This paper aims to find which transmission constraints could make MAS stable and obtain the consensus conditions for MAS \eqref{MAS_1}.

\subsection{Applications and Motivating Examples}\label{subsc:applications}
The distortion (attenuation or saturation) in information transmission or detection is an actual embodiment of transmission constraints. Those transmission constraints may be caused by objective physical constraints, or those constraints are added on purpose.

\subsubsection{Objective Constraints}
There are three kinds of objective constraints to show those transmission constraints are common in real world scenarios.
\begin{itemize}
\item \emph{Information distortion caused by transmission.}

Energy loss exists during the signal transmission, which may cause information distortion. The voltage drop on wires is an inevitable
phenomenon during signal transmission. If state of agent (or device) are represented by voltage of a signal, and this signal is transmitted on wires, then we can get an information distortion
\begin{equation*}
f_{ij}(x_i)=\frac{R_r}{R_L+R_r}x_i,
\end{equation*}
where $R_L$ is the resistance of wires, and $R_r$ is the equivalent resistance of the port.
\item \emph{Information distortion caused by detection.}

In real world scenarios, agents use sensors to get themselves or neighbors’ states. However, except for noise interference, state information cannot be obtained precisely due to the inherent characteristics of sensors. For example, temperature offset leads to signal fluctuation in ultrasonic distance measurements \cite{carullo2001ultrasonic}.
Likewise, the saturation characteristic of hall sensor may cause information distortion, i.e., $f_{ij}(x_i)=\text{sat}(x_i)$ \cite{ramsden2011hall}.
\item \emph{Information distortion caused by privacy protection.}

In social networks, individuals may express an opinion that is different from his/her private opinion, due to the pressure of conforming to a group standard or norm \cite{YE2019371,HOU2021125968}. Hence, $x_i$ could represent the private opinion, and $f_{ij}(x_i)$ is the expressed opinion.
\end{itemize}

\subsubsection{Subjective Constraints}
As a class of constraints, transmission constraints could make agent’s states converge into the expected set (see Theorem \ref{th:consensus} and Remark \ref{re:constrained state}). Especially, Since we do not specify the formula of transmission constraints, different transmission constraints can be designed to suit different scenarios, such as interval consensus \cite{fontan2019interval} and discarded consensus \cite{liu2012discarded}. Related discussions are in remarks \ref{re:interval consensus} and \ref{re:discarded consensus}.

Those above examples show that information distortion during transmission is a common phenomenon in the real world. Hence, study consensus under transmission constraints is necessary.

\subsection{Notations and Some Definitions}

\textbf{Notations:}
The set of positive integers is denoted by $\mathbb{N}^+$.
Consider a matrix $B=[b_{ij}]\in M_{m,n}$ and denote $|B|=[|b_{ij}|]$ (i.e., element-wise absolute value of matrix $B$).
$d^+Z(t)$ denotes the upper right Dini derivative of $Z(t)$.
The arrow `$\implies$' means `implies', and the arrow `$\iff$' means `if and only if'.
Denote sign function
\begin{equation*}
\text{sign}(x)=\begin{cases}
1, &\text{if } x>0,\\
0, &\text{if } x=0,\\
-1, &\text{if } x<0.
\end{cases}
\end{equation*}


The distance between interval $[ \partial_m, \partial_M]^n$ and vector $\mathbf{x}(t)$ is denoted by
\begin{align*}
\textit{distance}\big([ \partial_m, \partial_M]^n,\mathbf{x}(t)\big)=\min\limits_{c\in[ \partial_m, \partial_M]^n}\|\mathbf{x}(t)-c\|.
\end{align*}

Denote $\mathbf{e}=\{\mathbf{e}_1,\dots,\mathbf{e}_n\}^T$ to be an equilibrium of MAS \eqref{MAS_1}, then it can be concluded that for all $i\in\mathbf{N}$,
\begin{align*}
&\dot{x}_i(t)\big|_{\mathbf{e}_i}=\sum\limits_{j\in\mathcal{N}_i}a_{ij}\big(f_{ji}(\mathbf{e}_j)-\mathbf{e}_i\big)=0.
\end{align*}
Denote the error between the state $\mathbf{x}(t)$ and equilibrium $\mathbf{e}$ by $\varepsilon_i(t)=x_i(t)-\mathbf{e}_i$, $\forall\, i\in\mathbf{N}$.


Introduce the definition of consensus zone, and this work can be divided into two parts: the part of non-empty consensus zone and the part of empty consensus zone.

\begin{definition}
For MAS \eqref{MAS_1}, denote $\Theta_{ij}=\{x:f_{ij}(x)=x\}$, and
\textbf{consensus zone} $\Phi=\bigcap\limits_{(v_j,v_i)\in \mathcal{E}}\Theta_{ij}$.
Consensus zone means the transmission constraints vanish when all the states of MAS are in this consensus zone.
\end{definition}

Theorem \ref{th:consensus} shows that under the given conditions, MAS \eqref{MAS_1} reaches consensus and its states fall into consensus zone.

\begin{remark}
For any time, if an agent's state is in consensus zone, then the information it transmits to its neighbors is without transmission constraints. If all agents' initial states are in consensus zone, then the MAS becomes a standard consensus dynamics. In that case, under a strongly connected digraph (or the digraph has a spanning tree), the MAS will reach consensus and the consensus value is in the consensus zone. That is why we name it consensus zone.
\end{remark}

%

\section{Main Results}
In this section, we first analyze the convergence of MAS in subsection \ref{Part:CT_A}.
Secondly, for the non-empty consensus zone, we get the consensus conditions in subsection \ref{Part:CT_B}.
Thirdly, for the empty consensus zone, the system's states may achieve an equilibrium, and the existence, stability and uniqueness of equilibrium are studied in subsection \ref{Part:CT_C}.
\subsection{Convergence analysis}\label{Part:CT_A}

The following theorem states the conditions where the states of the multiagent system are bounded, and gives the boundary.
Furthermore, the conclusion of Theorem \ref{th:distance=0} plays an important role in the proofs of Theorems \ref{th:consensus} and \ref{th:exist_equilibrium}.

\begin{Theorem}\label{th:distance=0}
Along the system \eqref{MAS_1}, suppose Assumptions \ref{assumption_1}, \ref{assumption_2}, \ref{assumption_3} and \ref{assumption_4} hold, and $k_1k_2=1$.
Then for any MAS \eqref{MAS_1} and initial state $\mathbf{x}(t_0)\in\mathbb{R}^n$,
\begin{equation*}
\lim\limits_{t\to\infty}\textnormal{distance}\big([ \partial_m, \partial_M]^n,\mathbf{x}(t)\big)=0,
\end{equation*}
\textbf{if and only if} for any MAS \eqref{MAS_1}, initial state $\mathbf{x}(t_0)\in\mathbb{R}^n$, and $j\in\mathbf{N},\,i\in\mathcal{N}_j$, $\partial_m\le f_{ij}(x)\le\partial_M$, $x\in[\partial_m,\partial_M]$.
\end{Theorem}


\subsection{Nonempty Consensus Zone: Consensus}\label{Part:CT_B}
Then, we introduce conditions of transmission-constrained consensus.
The following theorem states the consensus conditions for MAS, and the condition about range of constraint functions is necessary and sufficient.

\begin{Theorem}\label{th:consensus}
Along the system \eqref{MAS_1}, suppose Assumptions \ref{assumption_1}, \ref{assumption_2}, \ref{assumption_3} and \ref{assumption_4} hold, and $f_{ij}(x)=x$, $x\in[\partial_m,\partial_M]$.
Then, for any MAS \eqref{MAS_1}, $i\in\mathbf{N}$, $\lim\limits_{t\to\infty}x_i(t)=v^*$, $v^*\in[\partial_m,\partial_M]$ \textbf{if and only if} for any MAS \eqref{MAS_1}, $k_1k_2\le1$.
\end{Theorem}

\begin{remark}\label{re:constrained state}
Theorem \ref{th:consensus} shows that although the initial state is not in the consensus zone, the transmission constraints will limit the final state of the system. Hence, it shows that imposing constraints on links of the interaction networks can indirectly limit agents' states.
\end{remark}

\begin{remark}
Under the conditions in Theorem \ref{th:consensus}, the constraint functions can be various functions. For example, constraint functions can be Sigmoid function or tanh function, which have many applications such as activation function in artificial neural networks, logistic function in biology, etc. More candidate constraint functions are given in the numeral example section.
\end{remark}

\begin{remark}\label{re:interval consensus}
In \cite{fontan2019interval}, the interval consensus problem is studied.
We propose a smooth interval consensus model:
\begin{equation*}
\dot{x}_i(t)=\sum_{j\in \mathcal{N}_i}a_{ij}\big[T_j\big(x_j(t)\big)-x_i(t)\big],\enspace i\in\mathbf{N},
\end{equation*}
\begin{equation*}
T(x)=\begin{cases}
\rho x+(1-\rho)q, &\text{if } x>q,\\
x, &\text{if } p\le x\le q,\\
\rho x+(1-\rho)p, &\text{if } x<p,
\end{cases}
\end{equation*}
where $\rho\in(0,1)$ is a constant. By Theorem \ref{th:consensus}, it can be concluded that the system will reach interval consensus.
\end{remark}

\begin{remark}\label{re:discarded consensus}
The discarded consensus problem is studied in \cite{liu2012discarded}, but the initial states of system must be in the constraint set. Another discarded consensus model can be proposed:
\begin{equation}
\dot{x}_i(t)=\sum_{j\in \mathcal{N}_i,x_j(t)\in\Omega_{c_i}}a_{ij}x_j(t)-\sum_{j\in \mathcal{N}_i}a_{ij}x_i(t),
\label{eq:discarded consensus}
\end{equation}
where $\Omega_{c_i}=[-c_i,c_i]$ is the constraint interval of agent $i$ \cite{liu2012discarded}.
By Theorem \ref{th:consensus}, we can get that the MAS \eqref{eq:discarded consensus} will reach discarded consensus with arbitrarily initial states.
\end{remark}

\begin{remark}
Consider the following multiagent system:
\begin{equation}\label{MAS_sin}
\dot{x}_i(t)=\sum_{j\in \mathcal{N}_i}a_{ij}\big[\sin\big(x_j(t)+\pi\big)-x_i(t)\big],\enspace i\in\mathbf{N}.
\end{equation}
Due to the consensus zone of MAS \eqref{MAS_sin} being $\Phi=\{0\}$, Theorem \ref{th:consensus} shows that the agents' states will converge to $0$, when the underlying directed graph $\mathcal{G}$ is strongly connected.
\end{remark}

\begin{Corollary}
Along the system \eqref{MAS_1}, suppose following conditions hold:
\begin{enumerate}
\item Assumption 2 holds;
\item the consensus zone $\Phi\ne\emptyset$;
\item for any $j\in\mathbf{N},\,i\in\mathcal{N}_j$ and $\omega\ne0$,
\begin{equation*}
-1<\frac{f_{ij}(x+\omega)-f_{ij}(x)}{\omega}\le1,\quad x\in\mathbb{R}.
\end{equation*}
\end{enumerate}
Then, $\forall\,i\in\mathbf{N}$, $\lim\limits_{t\to\infty}x_i(t)=v^*$, $v^*\in\Phi$.
\end{Corollary}

\subsection{Empty Consensus Zone: Existence, Stability and Uniqueness of Equilibria}\label{Part:CT_C}

In this part, the existence, stability and uniqueness of equilibria are discussed and proved.

Theorem \ref{th:exist_equilibrium} gives the existence conditions of equilibrium, which is a prerequisite for the Theorem \ref{th:single equilibrium}.

\begin{Theorem}\label{th:exist_equilibrium}
Suppose $\forall\,j\in\mathbf{N},\,i\in\mathcal{N}_j$, $f_{ij}(x)$ is a continuous function.
If there exists an interval $[\partial_m,\partial_M]$ and $\forall\,j\in\mathbf{N},\,i\in\mathcal{N}_j$,
\begin{equation*}
\partial_m\le f_{ij}(x)\le\partial_M,\quad x\in[\partial_m,\partial_M],
\end{equation*}
then the system \eqref{MAS_1} exists at least one equilibrium. In fact, all equilibria of the system lie within $[ \partial_m, \partial_M]^n$, if the following conditions hold:
\begin{enumerate}
\item Assumptions \ref{assumption_2}, \ref{assumption_3} and \ref{assumption_4} hold;
\item $k_{1}k_{2}=1$, and $\partial_m\le f_{ij}(x)\le\partial_M$, $x\in[\partial_m,\partial_M]$;
\item $f_{ij}(x)$ is a continuous function, $\forall\,j\in\mathbf{N},\,i\in\mathcal{N}_j$.
\end{enumerate}
\end{Theorem}


Theorem \ref{th:exist_equilibrium} establishes the existence of equilibrium and gives the region where all equilibria exist. But it does not illustrate whether the MAS will reach equilibria, not to mention the stability of equilibria.
Theorem \ref{th:converge_to_equilibrium} indicates that the system will converge to an asymptotically stable equilibrium, if some conditions hold.

\begin{Theorem}\label{th:converge_to_equilibrium}
Along the system \eqref{MAS_1}, suppose following conditions hold:
\begin{enumerate}
\item Assumption 2 holds;
\item there exists a equilibrium $\mathbf{e}=\{\mathbf{e}_1,\dots,\mathbf{e}_n\}^T$, i.e.,
\begin{equation*}
\dot{x}_i(t)\big|_{\mathbf{e}_i}=\sum\limits_{j\in\mathcal{N}_i}a_{ij}\big(f_{ji}(\mathbf{e}_j)-\mathbf{e}_i\big)=0,\enspace\forall\,i\in\mathbf{N}.
\end{equation*}
\item\label{item:th_converge_to_equilibrium_C3} there exist two rays
\begin{align*}
L_{e1}(\varepsilon)&=k_{e1}\varepsilon,\quad \varepsilon\in(-\infty,0];\\
L_{e2}(\varepsilon)&=k_{e2}\varepsilon,\quad \varepsilon\in[0,+\infty),
\end{align*}
where $k_{e1},k_{e2}<0$ and $k_{e1}k_{e2}=1$,
such that $\forall\,j\in\mathbf{N},\,i\in\mathcal{N}_j$,
\begin{align*}
\varepsilon\le f_{ij}(\mathbf{e}_i+\varepsilon)-f_{ij}(\mathbf{e}_i)< L_{e1}(\varepsilon),\quad \varepsilon\in(-\infty,0);\\
L_{e2}(\varepsilon)< f_{ij}(\mathbf{e}_i+\varepsilon)-f_{ij}(\mathbf{e}_i)\le \varepsilon,\quad\varepsilon\in(0,+\infty).
\end{align*}
\item for any $\varepsilon'\ne0$, there exist $j\in\mathbf{N},i\in\mathcal{N}_j$, such that $f_{ij}(\mathbf{e}_i+\varepsilon')$ is continuous on $\varepsilon'$ and $f_{ij}(\mathbf{e}_i+\varepsilon')-f_{ij}(\mathbf{e}_i)\ne \varepsilon'$.
\end{enumerate}
Then, the equilibrium $\mathbf{e}$ is unique and asymptotically stable, i.e., $\lim\limits_{t\to\infty}x_i(t)=\mathbf{e}_i$, $\forall\,i\in\mathbf{N}$.
\end{Theorem}

\begin{remark}
Unlike Theorem \ref{th:consensus}, the boundary rays in Theorem \ref{th:converge_to_equilibrium} are two clusters of parallel lines with the same slopes, but the endpoints may be different. The auxiliary lines in Fig. \ref{fig 0301} are the illustrations of two clusters of parallel lines.
\end{remark}

\begin{remark}
The unique equilibrium's values are only decided by the network structure and transmission constraint functions, but not related to the initial states of MAS \eqref{MAS_1}.
\end{remark}

The following theorem can be regarded as a combination of theorems \ref{th:exist_equilibrium} and \ref{th:converge_to_equilibrium}, which gives the conditions for the system to converge to an asymptotically stable equilibrium.

\begin{Theorem}\label{th:single equilibrium}
Along the system \eqref{MAS_1}, suppose following conditions hold:
\begin{enumerate}
\item Assumption 2 holds;
\item the consensus zone $\Phi=\emptyset$;
\item\label{item:th_single equilibrium_C3} for any $j\in\mathbf{N},\,i\in\mathcal{N}_j$ and $\omega\ne0$,
\begin{equation*}
-1<\frac{f_{ij}(x+\omega)-f_{ij}(x)}{\omega}<1,\quad x\notin\Theta_{ij}.
\end{equation*}
\end{enumerate}
Then there exists a unique, asymptotically stable equilibrium of the MAS \eqref{MAS_1}.
\end{Theorem}

\begin{remark}
Theorem \ref{th:converge_to_equilibrium} requires a known equilibrium of MAS.
In contrast, Theorem \ref{th:single equilibrium} relaxes the condition that the equilibrium is known, i.e., we just need to know the constraints functions and the connectivity of interaction networks, then we can predict the trajectory of MAS.
In conclusion, Theorem \ref{th:single equilibrium} states the existence, stability and uniqueness of Equilibria.
\end{remark}

\begin{Corollary}\label{co:single equilibrium}
Suppose the directed graph $\mathcal{G}$ is strongly connected and the consensus zone $\Phi=\emptyset$.
If for all $j\in\mathbf{N},\,i\in\mathcal{N}_j$, $f_{ij}(x_i)=k_{ij}(x_i)x_i+m_{ij}(x_i)$ is a continuous and piecewise linear function with its slopes $k_{ij}\in(-1,1]$ and $m_{ij}=0$ when $k_{ij}=1$, then the MAS \eqref{MAS_1} has a unique, asymptotically stable equilibrium point.
\end{Corollary}

\section{Numeral Example}
In this section, four numeral examples are presented to illustrate the theorems and corollaries proposed in this paper.
Examples \ref{ex1} and \ref{ex2} illustrate the consensus theorem.
Additionally, example \ref{ex3} illustrates the theorem for stability, uniqueness of equilibrium, i.e., Theorem \ref{th:single equilibrium}.

In all following examples, the number of agents $n=5$ with underlying strongly connected graphs.
For simplicity, in examples \ref{ex2} and \ref{ex3}, let $f_{ij}(x)=f_i(x)$, $\forall\,j\in\mathbf{N}$.

\begin{example}\label{ex1}
\begin{figure}[t]
\centering
\includegraphics[scale=0.5]{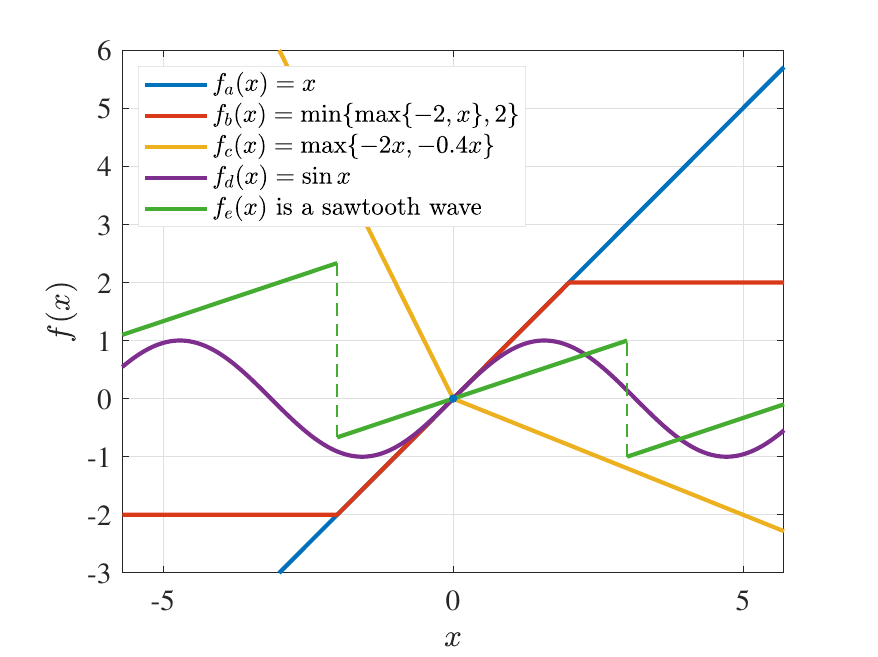}
\caption{The constraint functions in Example \ref{ex1}.}
\label{fig 0101}
\end{figure}
\begin{figure}[t]
\centering
\includegraphics[scale=0.5]{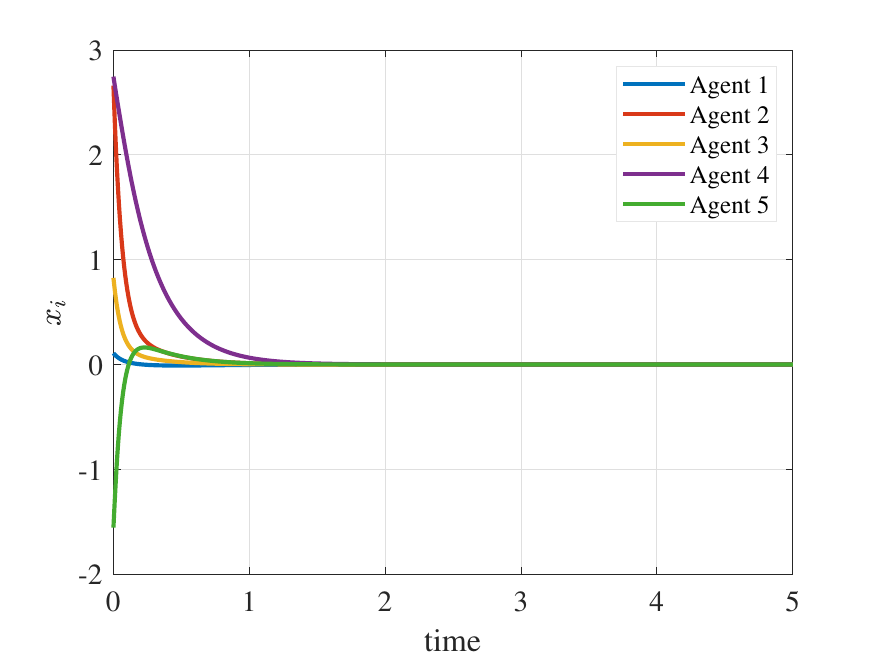}
\caption{The trajectories of $\mathbf{x}(t)$ in Example \ref{ex1}.}
\label{fig 0102}
\end{figure}
The adjacency matrix in this example is
\begin{equation*}
\mathcal{A}_1 = \begin{bmatrix}
0 		& 		0 			& 3.6 		& 0 		& 0		\\
0 		& 		0 			& 4.6		& 1.3 	& 6.5	\\
3.6 	& 		0 			& 0 			& 0 		& 7.6	\\
0.5 	& 		1.4 		& 2.1 		& 0 		& 0		\\
2.9 	& 		6.5 		& 0 			& 0 		& 0
\end{bmatrix},
\end{equation*}
Fig. \ref{fig 0101} shows candidates for the constraint function imposed on the information transmissions, and the configuration of constraint functions is shown in Table \ref{table2}. It shows that the consensus zone $\Phi=\{0\}$. Fig. \ref{fig 0102} shows that MAS achieves transmission-constrained consensus with $\lim\limits_{t\to\infty}x(t)=0$.
In this example, the constraint function $f_e(x)$ is a piecewise continuous function similar to a sawtooth wave.
And the constraint function $f_c(x)$ can be chosen approximately as the boundary rays since it satisfies the Condition (ii) of Theorem \ref{th:consensus} and $k_1k_2=0.8<1$.

\renewcommand\arraystretch{1.3}
\begin{table}[t]
\caption{Configuration of transmission constraints in Example \ref{ex1}}
\label{table2}
\centering
\begin{tabular}{|c|c|c|c|c|c|}
\hline
\multicolumn{6}{|c|}{Transmission Constraints $f_{ij}$} \\
\hline
\diagbox{$i$}{$f_{ij}$}{$j$} & 1 & 2 & 3 & 4 & 5\\
\hline
1 & \text{---} & \text{---} & $(f_b+f_d)/2$ & \text{---} & \text{---} \\
\hline
2 & \text{---} & \text{---} & $f_a$ & $f_b$ & $(f_e+f_a)/2$ \\
\hline
3 & $f_d$ & \text{---} & \text{---} & \text{---} & $f_c$ \\
\hline
4 & $f_a$ & $f_b$ & $(f_c+f_d)/2$ & \text{---} & \text{---} \\
\hline
5 & $f_c$ & $f_e$ & \text{---} & \text{---} & \text{---} \\
\hline
\end{tabular}
\end{table}
\renewcommand\arraystretch{1}
\end{example}

\begin{example}\label{ex2}
\begin{figure}[t]
\centering
\includegraphics[scale=0.5]{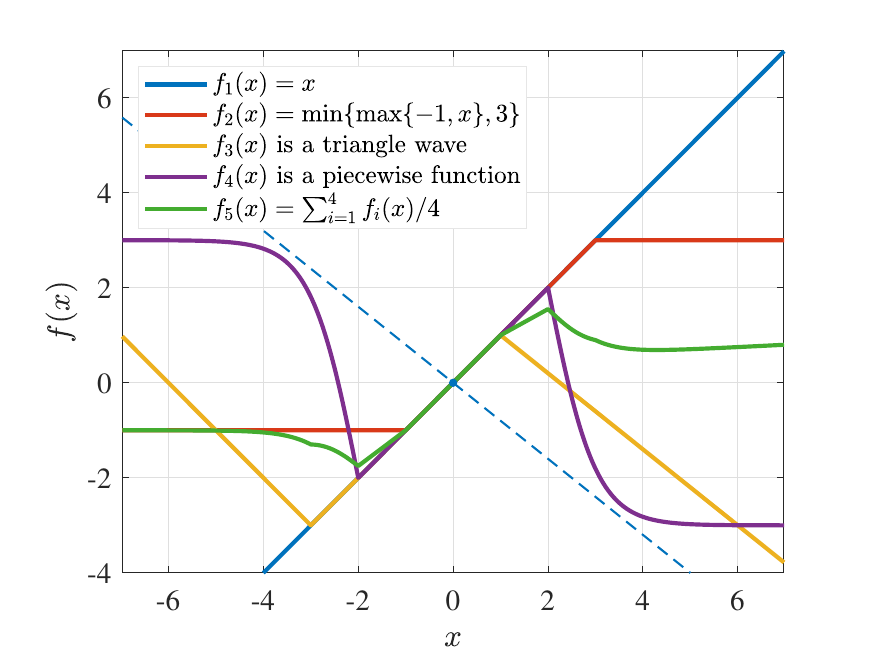}
\caption{The constraint functions in Example \ref{ex2}.}
\label{fig 0201}
\end{figure}
\begin{figure}[t]
\centering
\includegraphics[scale=0.5]{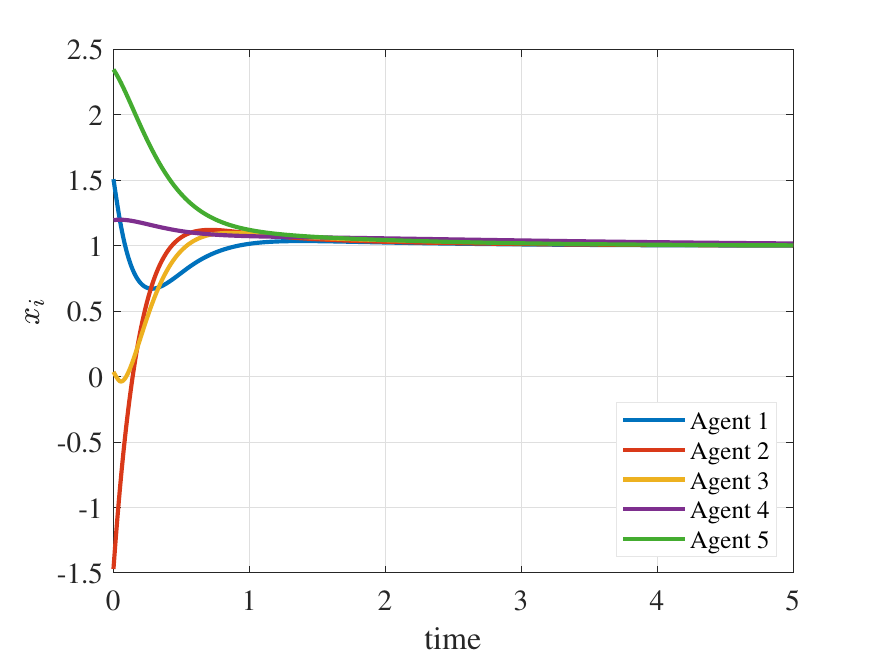}
\caption{The trajectories of $\mathbf{x}(t)$ in Example \ref{ex2}.}
\label{fig 0202}
\end{figure}
The adjacency matrix in this example is
\begin{equation*}
\mathcal{A}_2 = \begin{bmatrix}
0 		& 		2.5 			& 0.6 		& 0 		& 0		\\
0 		& 		0 			& 0		& 0 	& 4.5	\\
0 	& 		5.6 			& 0 			& 3.3 		& 0	\\
0.5 	& 		0 		& 0 		& 0 		& 0		\\
1.9 	& 		0 		& 0 			& 0 		& 0
\end{bmatrix}.
\end{equation*}
In Fig. \ref{fig 0201}, we can get that the consensus zone $\Phi=[-1,1]^5$.
The auxiliary line in Fig. \ref{fig 0201} represents the boundary rays with $k_1k_2=0.64<1$.
Fig. \ref{fig 0202} shows that $x(t)\to\Phi$ as $t\to\infty$.
\end{example}

\begin{example}\label{ex3}
\begin{figure}[t]
\centering
\includegraphics[scale=0.5]{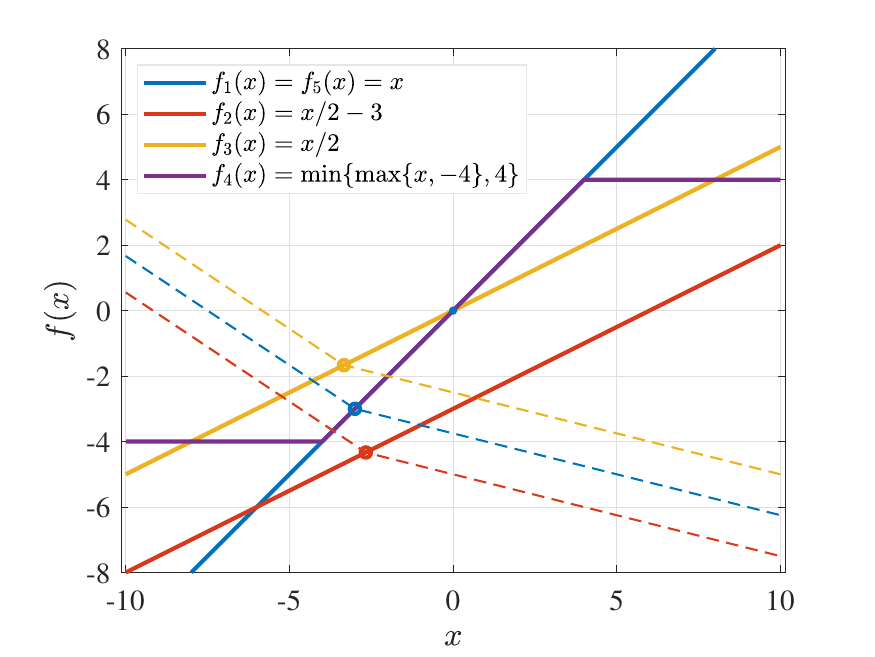}
\caption{The constraint functions in Example \ref{ex3}.}
\label{fig 0301}
\end{figure}
\begin{figure}[t]
\centering
\includegraphics[scale=0.5]{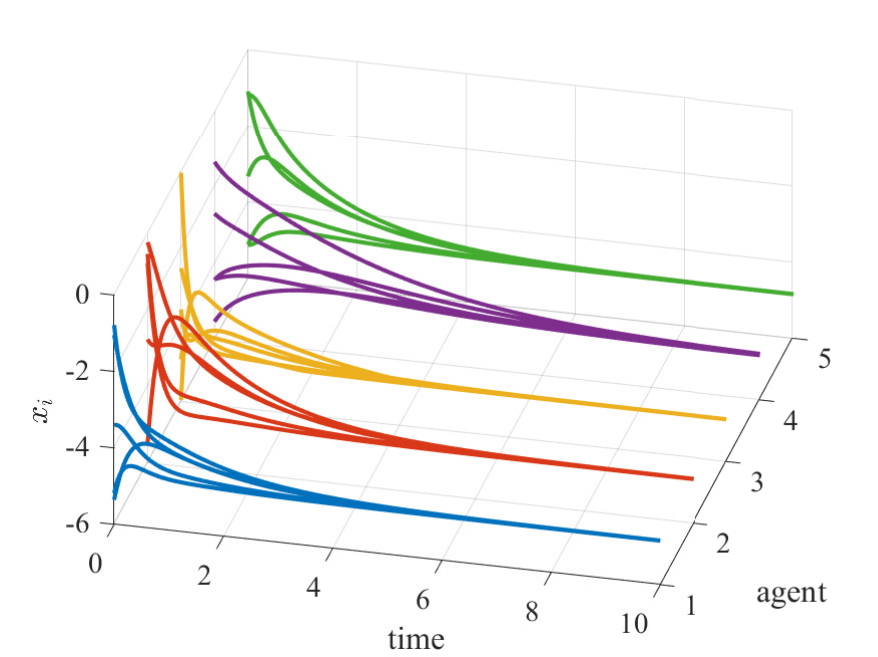}
\caption{The trajectories of $\mathbf{x}(t)$ in Example \ref{ex3}.}
\label{fig 0302}
\end{figure}
The adjacency matrix in this continuous-time example is $\mathcal{A}_2$.
Fig. \ref{fig 0302} shows that no matter which initial states of agents are, MAS will reach the same equilibrium, i.e., unique equilibrium (see the points $\big(\mathbf{e}_i,f_i(\mathbf{e}_i)\big)$ shown by circles in Fig. \ref{fig 0301}).
It is easy to know that for any equilibrium $\mathbf{e}$, there exist two clusters of rays (see auxiliary lines in Fig. \ref{fig 0301}) satisfying the Condition (iii) of Theorem \ref{th:converge_to_equilibrium} or the Condition (iii) of Theorem \ref{th:single equilibrium}, which means that the system will converge to a unique, asymptotically stable equilibrium, even though we do not know the value of equilibrium.
\end{example}


\section{Conclusion}
This paper focuses on the transmission-constrained consensus problem of multiagent networks, where information transmissions between agents are affected by irregular constraint functions. We obtain the necessary and sufficient conditions about the range of transmission constraint functions where agents' states can converge to consensus.
Due to the piecewise continuous constraint functions, the LaSalle invariance principle is not applicable in those proofs. We construct a sophisticated Lyapunov function and discuss the boundaries of multiple limit points of MAS states to facilitate the convergence analysis.
Meanwhile, in some cases where the system cannot achieve consensus, there is an asymptotically stable  and unique equilibrium independent of the initial values of agents' states. Finally, the numerical simulations are presented to verify the effectiveness of theoretical results.


\appendices
\section{Proof of Theorem \ref{th:distance=0}}
\subsection{Technical lemmas}
First of all, we introduce some technical lemmas.
\begin{Lemma}\label{le:der_to_max}
\emph{\bf{(Lemma 2.2 in \cite{lin2007state})}} If for all $i\in\mathbf{N}$, $Z_i(\mathbf{x}):\mathbb{R}^d\to\mathbb{R}$ is of class $C^1$, and denote $Z(\mathbf{y})=\max\limits_{i\in\mathbf{N}}Z_i(\mathbf{y})$. Denote $\mathsf{N}_m(t)=\{i\in\mathbf{N}:Z(\mathbf{y})=Z_i(\mathbf{y})\}$ the indices set in which the maximum is reached at time $t$. Then it turns out that $d^+Z\big(\mathbf{y}(t)\big)=\max\limits_{i\in\mathsf{N}_m(t)}\dot{Z}_i\big(\mathbf{y}(t)\big)$.
\end{Lemma}

%



\begin{Lemma}\label{le:one_line_crosses_boundary}
If $\partial\in[\partial_m,\partial_M]$ and $k_1k_2=1$, then
\begin{equation*}
\partial_M-\partial_m\ge\min\{(1-k_2)(\partial_M-\partial),(1-k_1)(\partial-\partial_m)\}.
\end{equation*}
\end{Lemma}

\begin{proof}
When $\partial_M-\partial_m=0$ or $(\partial_M-\partial)(\partial-\partial_m)=0$, the conclusion is obvious.

When $\partial_M-\partial_m>0$ and $(\partial_M-\partial)(\partial-\partial_m)>0$, we use a contradiction argument to prove it.
Let $\partial_M-\partial_m<\min\{(1-k_2)(\partial_M-\partial),(1-k_1)(\partial-\partial_m)\}$, i.e.,
\begin{equation}\label{eq:le_one_line_crosses_boundary_1}
\left\{
\begin{aligned}
\partial_M-\partial_m&<(1-k_2)(\partial_M-\partial);\\
\partial_M-\partial_m&<(1-k_1)(\partial-\partial_m).
\end{aligned}
\right.
\end{equation}
Since $\partial\in(\partial_m,\partial_M)$, we let $\partial=\rho\partial_M+(1-\rho)\partial_m$ with $\rho\in(0,1)$. Then, \eqref{eq:le_one_line_crosses_boundary_1} can be rewritten as
\begin{equation*}
\left\{
\begin{aligned}
\partial_M-\partial_m&<(1-k_2)(1-\rho)(\partial_M-\partial_m);\\
\partial_M-\partial_m&<(1-k_1)\rho(\partial_M-\partial_m).
\end{aligned}
\right.
\end{equation*}
Then, we can get that $-k_2>\frac{\rho}{1-\rho}>0$ and $-k_1>\frac{1-\rho}{\rho}>0$, which implies that $k_1k_2>1$. We get a contradiction, and prove the Lemma \ref{le:one_line_crosses_boundary}.
\end{proof}

The following lemma shows the implicit inequality from the given condition, and it helps us discuss the boundedness of transmission-constrained consensus dynamics.

\begin{Lemma}\label{le:x_M and x_m}
Denote $x_m(t)=\min\limits_{i\in\mathbf{N}}x_i(t)$, $x_M(t)=\max\limits_{i\in\mathbf{N}}x_i(t)$, $L_1\big(x_i(t)\big)=k_1x_i(t)+(1-k_1)\partial$, $L_2\big(x_i(t)\big)= k_2x_i(t)+(1-k_2)\partial$ and
\begin{align*}
Y(t)=\max\big\{L_1\big(x_m(t)\big)-x_m(t),\;x_M(t)-L_2\big(x_M(t)\big)\big\},
\end{align*}
where $\partial$ is a constant value.
If $k_1,k_2<0$ and $k_1k_2=1$, then $\forall\,i\in\mathbf{N}$,
\begin{enumerate}
\item\label{item:con1} $Y(t)=x_M(t)-L_2\big(x_M(t)\big)\implies$$x_M(t)\ge L_1\big(x_i(t)\big)$;
\item\label{item:con2} $Y(t)=L_1\big(x_m(t)\big)-x_m(t)\implies$$x_m(t)\le L_2\big(x_i(t)\big)$.
\end{enumerate}
\end{Lemma}
\begin{proof}
We first discuss Item \ref{item:con1}, i.e., the case where $Y(t)=x_M(t)-L_2\big(x_M(t)\big)$, which means that $x_M(t)\ge\partial$.

Note that for any $i\in\mathbf{N}$, if $x_i(t)\ge\partial$, it is easy to know that,
\begin{equation*}
x_M(t)\ge\partial\ge\partial+(1-k_1)\big(\partial-x_i(t)\big)=L_1(x_i(t)).
\end{equation*}

For any $i\in\mathbf{N}$, if $x_i(t)<\partial$,
\begin{align*}
&Y(t)=x_M(t)-L_2\big(x_M(t)\big)\\
\iff& (1-k_2) \big(x_M(t)-\partial\big)\ge (1-k_1)\big(\partial-x_i(t)\big)\\
\iff& x_M(t)-\partial\ge \frac{1-k_1}{1-k_2}\big(\partial-x_i(t)\big)\\
\iff& x_M(t)-\partial\ge k_1(x_i-\partial).
\end{align*}

Then we can get that $x_M(t)-k_1x_i(t)-(1-k_1)\partial\ge0$, which implies
\begin{equation*}
x_M(t)\ge L_1\big(x_i(t)\big)=k_1x_i(t)+(1-k_1)\partial,\enspace \forall\,i\in\mathbf{N}.
\end{equation*}

Therefore, the proof of Item \ref{item:con1} is completed.
The proof method of Item \ref{item:con2} is similar to that of Item \ref{item:con1}, and hence is omitted here.
Hence, Lemma \ref{le:x_M and x_m} is proved.
\end{proof}

\begin{Lemma}\label{le:positive_invariant}
For the MAS \eqref{MAS_1}, if there exists an interval $[ \partial_m, \partial_M]^n$ such that for all $j\in\mathbf{N},\,i\in\mathcal{N}_j$,
\begin{equation*}
\partial_m\le f_{ij}(x)\le\partial_M,\enspace x\in[\partial_m,\partial_M],
\end{equation*}
then $[ \partial_m, \partial_M]^n$ is a positively invariant set.
\end{Lemma}

\begin{proof}
The dynamics of MAS \eqref{MAS_1} can be rewritten as
\begin{equation}\label{eq:vector field}
\dot{x}(t)=h\big(x(t)\big)=\Big(h_1\big(x(t)\big),\dots, h_n\big(x(t)\big)\Big)^T,
\end{equation}
where $h_i\big(x(t)\big)=\sum_{j\in \mathcal{N}_i}a_{ij}\Big[f_{ji}\big(x_j(t)\big)-x_i(t)\Big]$.

The initial states of \eqref{eq:vector field} is $x_0=x(t_0)$. Assume that $x(t_0)\in[ \partial_m, \partial_M]^n$. Since the vector field $h$ is pointing inwards $[ \partial_m, \partial_M]^n$ that is an $n$-dimensional cube, it concludes that
\begin{equation*}
x(t)\in[ \partial_m, \partial_M]^n,\enspace\forall\,t\ge t_0.
\end{equation*}

It shows that $[ \partial_m, \partial_M]^n$ is a positively invariant set and the proof is completed.
\end{proof}

\begin{Lemma}\label{le:non-increasing function}
Along the system \eqref{MAS_1}, suppose there exist an interval $[\partial_m,\partial_M]$, a value $\partial\in[\partial_m,\partial_M]$ and two rays
\begin{align*}
L_{1}(x)&=k_{1}(x-\partial)+\partial,\quad x\in(-\infty,\partial];\\
L_{2}(x)&=k_{2}(x-\partial)+\partial,\quad x\in[\partial,+\infty),
\end{align*}
where $k_{1}k_{2}=1$,
such that $\forall\,j\in\mathbf{N},\,i\in\mathcal{N}_j$,
\begin{equation*}
\begin{alignedat}{2}
&x\le f_{ij}(x)\le L_{1}(x),\quad & &x\in(-\infty,\partial_m);\\
&\partial_m\le f_{ij}(x)\le\partial_M,\quad & &x\in[\partial_m,\partial_M];\\
&L_{2}(x)\le f_{ij}(x)\le x, & &x\in(\partial_M,+\infty).
\end{alignedat}
\end{equation*}
Denote $x_m(t)=\min\limits_{i\in\mathbf{N}}x_i(t)$, $x_M(t)=\max\limits_{i\in\mathbf{N}}x_i(t)$ and
\begin{align*}
Y(t)=&\max\big\{\partial_M-\partial_m,x_M(t)-\partial_m,\partial_M-x_m(t),\\
&x_M(t)-L_2\big(x_M(t)\big),L_1\big(x_m(t)\big)-x_m(t)\big\}.
\end{align*}
If $\partial_M-\partial_m\ge\max\{(1-k_2)(\partial_M-\partial),(1-k_1)(\partial-\partial_m)\}$, then $Y(t)$ is a non-increasing function for any initial state $x_*\in\mathbb{R}^n$.
\end{Lemma}

\begin{proof}
By the structure of $Y(t)$, there exists five cases:
\begin{enumerate}
\item\label{item:th_distance=0_partA_case1} $Y(t)=Y_1(t)=x_M(t)-L_2\big(x_M(t)\big)$;
\item\label{item:th_distance=0_partA_case2} $Y(t)=Y_2(t)=x_M(t)-\partial_m$;
\item\label{item:th_distance=0_partA_case3} $Y(t)=Y_3(t)=L_1\big(x_m(t)\big)-x_m(t)$;
\item\label{item:th_distance=0_partA_case4} $Y(t)=Y_4(t)=\partial_M-x_m(t)$;
\item\label{item:th_distance=0_partA_case5} $Y(t)=Y_5(t)=\partial_M-\partial_m$.
\end{enumerate}

The Case \ref{item:th_distance=0_partA_case1} is analyzed firstly.
Since
\begin{equation*}
x_M(t)-L_2\big(x_M(t)\big)\ge\partial_M-\partial_m\ge(1-k_2)(\partial_M-\partial),
\end{equation*}
it implies that $x_M(t)\ge\partial_M$.

Denote $\mathcal{I}_M(t)=\big\{k:x_k(t)=\max\limits_{l\in\mathbf{N}}x_l(t)\big\}$. By Lemma \ref{le:der_to_max}, we have
\begin{align*}
&\phantom{=}d^+Y_1(t)=d^+\max\limits_{i\in\mathbf{N}}\Big\{(1-k_2)\big(x_i(t)-\partial\big)\Big\}\\
&=\max\limits_{i\in\mathcal{I}_M(t)}\Big\{(1-k_2)\sum\limits_{j\in\mathcal{N}_{i}(t)}a_{ij}\Big(f_{ji}\big(x_j(t)\big)-x_{i}(t)\Big)\Big\}.
\end{align*}

For all $i\in\mathcal{I}_M(t)$, which means that $x_i(t)=x_M(t)$, and we conduct the following analysis:
\begin{equation*}
\begin{cases}
x_M\ge L_{1}(x_j)> f_{ji}(x_j) &\text{if } x_j\le\partial_m;\\
x_M\ge\partial_M\ge f_{ji}(x_j) &\text{if } x_j\in[\partial_m, \partial_M];\\
x_M\ge x_j\ge f_{ji}(x_j) &\text{if } x_j>\partial_M,
\end{cases}
\end{equation*}
where the first inequality follows from Lemma \ref{le:x_M and x_m} and the fact that $x_M(t)-L_2\big(x_M(t)\big)\ge L_1\big(x_m(t)\big)-x_m(t)$.

Then, it can be concluded that $d^+Y_1(t)\le0$ when $Y(t)=Y_1(t)=x_M(t)-L_2\big(x_M(t)\big)$.

Secondly, the Case \ref{item:th_distance=0_partA_case2} is discussed. Since $x_M(t)-\partial_m\ge\partial_M-\partial_m$, it implies that $x_M(t)\ge\partial_M$.

Since $x_M(t)-\partial_m\ge L_1\big(x_m(t)\big)-x_m(t)$, we can get that
\begin{align*}
x_M(t)\ge& k_1x_m(t)+(1-k_1)\partial+\partial_m-x_m(t)\\
=&L_{1}\big(x_m(t)\big)+\partial_m-x_m(t)\\
\ge&L_{1}\big(x_i(t)\big)+\partial_m-x_m(t),\enspace\forall\,i\in\mathbf{N}.
\end{align*}

It turns out that
\begin{equation*}
\begin{cases}
x_M\ge L_{1}(x_j)\ge f_{ji}(x_j) &\text{if } x_j\le\partial_m;\\
x_M\ge\partial_M\ge f_{ji}(x_j) &\text{if } x_j\in[\partial_m, \partial_M];\\
x_M\ge x_j\ge f_{ji}(x_j) &\text{if } x_j>\partial_M,
\end{cases}
\end{equation*}
where the first inequality follows from $x_M\ge L_{1}(x_i)+\partial_m-x_m\ge L_{1}(x_i)$ when $x_m\le\partial_m$.

Then, it shows that
\begin{align*}
\phantom{=}d^+Y_2(t)=\max\limits_{i\in\mathcal{I}_M(t)}\Big\{\sum\limits_{j\in\mathcal{N}_{i}(t)}a_{ij}\Big(f_{ji}\big(x_j(t)\big)-x_{i}(t)\Big)\Big\}\le0.
\end{align*}

The analyses of cases \ref{item:th_distance=0_partA_case3} and \ref{item:th_distance=0_partA_case4} are symmetric to those of cases \ref{item:th_distance=0_partA_case1} and \ref{item:th_distance=0_partA_case2}, hence they are omitted. As for Case \ref{item:th_distance=0_partA_case5}, the conclusion is obvious.

Therefore, by the above five cases, it can be concluded that $Y(t)$ is a non-increasing function.
\end{proof}

\begin{Lemma}\label{le:the boundary of x_M and x_m}
Suppose the MAS \eqref{MAS_1} satisfies the conditions in Lemma \ref{le:non-increasing function}. Denote the initial time $t_0$. Then for any $t\ge t_0$, we have
\begin{enumerate}
\item $Y(t_0)=\partial_M-\partial_m\implies \partial_m\le x_i(t)\le\partial_M,\,\forall\,i\in\mathbf{N}$;
\item $Y(t_0)=x_M(t_0)-L_2\big(x_M(t_0)\big)$$\implies$$ x_m(t)\ge L_{2}\big(x_M(t_0)\big)$;
\item $Y(t_0)=L_1\big(x_m(t_0)\big)-x_m(t_0)$$\implies$$ x_M(t)\le L_{1}\big(x_m(t_0)\big)$;
\item $Y(t_0)=x_M(t_0)-\partial_m$$\implies$$ x_m(t)\ge\min\{x_m(t_0),\partial_m\}$;
\item $Y(t_0)=\partial_M-x_m(t_0)$$\implies$$ x_M(t)\le\max\{x_M(t_0),\partial_M\}$.
\end{enumerate}
\end{Lemma}

\begin{proof}
$Y(t_0)=\partial_M-\partial_m$ means that $\mathbf{x}(t_0)\in[\partial_m,\partial_M]^n$. By Lemma \ref{le:positive_invariant}, we can get that $[\partial_m,\partial_M]^n$ is a positively invariant set, and this case is proven trivially.

When $Y(t_0)=x_M(t_0)-L_2\big(x_M(t_0)\big)$, if $ x_m(t)< L_{2}\big(x_M(t_0)\big)$, we have
\begin{align*}
&Y(t)\ge L_1\big(x_m(t)\big)-x_m(t)> (1-k_1)\big(\partial-L_2(x_M(t_0))\big)\\
=& -k_2(1-k_1)\big(x_M(t_0)-\partial\big)= (1-k_2)\big(x_M(t_0)-\partial\big)\\
=&Y(t_0),
\end{align*}
which contradicts the Lemma \ref{le:non-increasing function}. By symmetry, the case where $Y(t_0)=L_1\big(x_m(t_0)\big)-x_m(t_0)$ is also proven.

When $Y(t_0)=x_M(t_0)-\partial_m$, it means that $L_{2}\big(x_M(t_0)\big)\ge \partial_m$. Since $x_M(t)\le x_M(t_0)$, $\forall\,t\ge t_0$, we have $L_{2}\big(x_i(t)\big)\ge L_{2}\big(x_M(t_0)\big)$, $\forall\,i\in\mathbf{N},\,t\ge t_0$. Hence, it can be concluded that $f_{ji}\big(x_j(t)\big)\ge\min\{x_j(t),\partial_m,L_{2}\big(x_j(t)\big)\}\ge\min\{x_m(t),\partial_m\}$, $\forall\,i,j\in\mathbf{N}$. Therefore, we can get $ x_m(t)\ge\min\{x_m(t_0),\partial_m\}$. By symmetry, the case where $Y(t_0)=\partial_M-x_m(t_0)$ is also proven.
This proof is completed.

When $Y(t_0)=\partial_M-x_m(t_0)$, it shows that $L_1\big(x_m(t_0)\big)\le\partial_M$. Since $x_m(t)\ge x_m(t_0)$, $\forall\,t\ge t_0$. Hence, it can be concluded that $f_{ji}\big(x_j(t)\big)\le \max\{x_j(t),\partial_M,L_1\big(x_m(t)\big)\}\le \max\{x_M(t),\partial_M\}$, $\forall\,i,j\in\mathbf{N}$. Then, a contradiction argument is used to prove that $x_M(t)\le \max\{x_M(t_0),\partial_M\}$, $\forall\,t\ge t_0$. Assume that $\exists\,t_*\ge t_0$, $x_M(t_*)>\max\{x_M(t_0),\partial_M\}$. Hence, there exists a $T\ge t_0$, such that for any $t\in[t_0,T]$,
\begin{equation*}
\left\{
\begin{aligned}
&x_M(t)\le\max\{x_M(t_0),\partial_M\},\\
&x_M(T)=\max\{x_M(t_0),\partial_M\},\\
&d^+x_M(T)>0.
\end{aligned}
\right.
\end{equation*}
Since
\begin{align*}
&d^+x_M(T)=\max\limits_{i\in\mathcal{I}_M(T)}\dot{x}_i(T)\\
=&\max\limits_{i\in\mathcal{I}_M(T)}\sum\limits_{j\in\mathcal{N}_{i}(T)}a_{ij}\Big(f_{ji}\big(x_j(T)\big)-x_{i}(T)\Big)\\
\le&\max\limits_{i\in\mathcal{I}_M(T)}\sum\limits_{j\in\mathcal{N}_{i}(T)}a_{ij}\big(\max\{x_i(T),\partial_M\}-x_{i}(T)\big)\\
=&\max\limits_{i\in\mathcal{I}_M(T)}\sum\limits_{j\in\mathcal{N}_{i}(T)}a_{ij}\big(\max\{x_M(t_0),\partial_M\}\\
&-\max\{x_M(t_0),\partial_M\}\big)=0,
\end{align*}
which leads to a contradiction and it shows that $x_M(t)\le \max\{x_M(t_0),\partial_M\}$, $\forall\,t\ge t_0$. By symmetry, the case where $Y(t_0)=\partial_M-x_m(t_0)$ is also proven.
\end{proof}

\subsection{Proof of Theorem \ref{th:distance=0}}
\begin{proof}
\subsubsection{Necessity}
A contradiction argument is applied to prove the necessity.

For simplicity, we assume that there are only two agents in MAS \eqref{MAS_1}, i.e., $\mathbf{N}=\{1,2\}$.
Since $\mathcal{G}$ is strongly connected, it turns out that $\mathcal{N}_1=\{2\}$ and $\mathcal{N}_2=\{1\}$.

Denote the initial time $t_0\ge0$. Suppose there exist $j\in\mathbf{N},i\in\mathcal{N}_j$ and $x_j(t_0)\in[\partial_m, \partial_M]$ such that
\begin{equation*}
f_{ji}\big(x_j(t_0)\big)=\partial_M+\omega,\enspace\omega>0.
\end{equation*}

Without loss of generality, assume that $j=1$ and $i=2$, i.e., $f_{12}\big(x_1(t_0)\big)=\partial_M+\omega$ in which $x_1(t_0)\in[\partial_m, \partial_M]$.
Let $x_2(t_0)=f_{12}\big(x_1(t_0)\big)$ and $f_{21}\big(x_2(t_0)\big)=x_1(t_0)$, then it can be concluded that for all $t\ge t_0$,
\begin{align*}
\dot{x}_1(t)&=a_{12}\Big(f_{21}\big(x_2(t)\big)-x_1(t)\Big)=0,\\
\dot{x}_2(t)&=a_{21}\Big(f_{12}\big(x_1(t)\big)-x_2(t)\Big)=0.
\end{align*}
which implies that $x_1(t)=x_1(t_0)$, $x_2(t)=x_2(t_0)$, $\forall\,t\ge t_0$.

Moreover, because
\begin{equation*}
\left\{
\begin{aligned}
&x_1(t_0)\in[\partial_m, \partial_M],\\
&f_{12}\big(x_1(t_0)\big)=\partial_M+\omega>\partial_M,\\
&x_2(t_0)=f_{12}\big(x_1(t_0)\big)=\partial_M+\omega>\partial_M,\\
&f_{21}\big(x_2(t_0)\big)=x_1(t_0)\le\partial_M,
\end{aligned}
\right.
\end{equation*}
it is easy to find two rays $L_{1}$ and $L_{2}$ satisfying the Assumptions \ref{assumption_3}, \ref{assumption_4}, and the condition $k_1k_2=1$ is also satisfied.

Since $\forall\,t\ge t_0$, $x_2(t)=\partial_M+\omega>\partial_M$, it shows that
\begin{equation*}
\lim\limits_{t\to\infty}\textnormal{distance}\big([ \partial_m, \partial_M]^n,\mathbf{x}(t)\big)\ne0.
\end{equation*}

Hence, we get a contradiction and the proof for the necessity statement of Theorem \ref{th:distance=0} is proved.

\subsubsection{Sufficiency}
We prove it in three steps.
\begin{step1}\label{th_distance=0_partA_step1}
Since $\partial\in[\partial_m,\partial_M]$ and $k_1k_2<1$, by Lemma \ref{le:one_line_crosses_boundary}, there exist two possibilities:
\begin{enumerate}
\item\label{item:th_distance=0_P1} $\partial_M-\partial_m\ge\max\{(1-k_2)(\partial_M-\partial),(1-k_1)(\partial-\partial_m)\}$;
\item\label{item:th_distance=0_P2} $\partial_M-\partial_m<(1-k_2)(\partial_M-\partial)$ or $\partial_M-\partial_m<(1-k_1)(\partial-\partial_m)$.
\end{enumerate}

We discuss the Possibility \ref{item:th_distance=0_P1}) in the rest of Step \ref{th_distance=0_partA_step1}, and the Possibility \ref{item:th_distance=0_P2}) is analyzed in Step \ref{th_distance=0_partB_step1}.

Assume that $\partial_M-\partial_m\ge\max\{(1-k_2)(\partial_M-\partial),(1-k_1)(\partial-\partial_m)\}$. Form Lemma \ref{le:non-increasing function}, we have that $Y(t)$ is a non-increasing function.

Let $n$ be the number of agents. Since
\begin{align*}
Y(t)=&\max\big\{\partial_M-\partial_m,x_M(t)-\partial_m,\partial_M-x_m(t),\\
&x_M(t)-L_2\big(x_M(t)\big),L_1\big(x_m(t)\big)-x_m(t)\big\},
\end{align*}
we continue this proof case by case.

\begin{case}\label{case:Y_1}
$Y(t_0)=Y_1(t_0)=x_M(t_0)-L_2\big(x_M(t_0)\big)$.

By Lemma \ref{le:the boundary of x_M and x_m}, we have for any $t\ge t_0$,
\begin{align*}
&L_{1}\big(x_m(t)\big)=k_{1}\big(x_m(t)-\partial\big)+\partial\\
\le& k_{1}k_{2}x_M(t_0)+k_{1}(1-k_{2})\partial+(1-k_{1})\partial=x_M(t_0).
\end{align*}

Choose $i_0\in\mathcal{I}_0:=\{i:x_i(t_0)=x_m(t_0)\}$. For any $j\in\mathcal{N}_{i_0}$, we can get that
\begin{equation*}
f_{ji_0}\big(x(t)\big)\le\max\big\{x_M(t),L_{1}\big(x_m(t)\big),\partial_M\big\}\le x_M(t_0).
\end{equation*}

It implies that
\begin{align*}
\dot{x}_{i_0}(t)&=\sum_{j\in\mathcal{N}_{i_0}}a_{i_0j}\big[f_{ji_0}\big(x_j(t)\big)-x_{i_0}(t)\big]\\
&\le \alpha_{i_0}[x_M(t_0)-x_{i_0}(t)],
\end{align*}
which implies that
\begin{equation*}
x_{i_0}(t)\le e^{-\alpha_{i_0}(t-t_0)}x_m(t_0)+[1-e^{-\alpha_{i_0}(t-t_0)}]x_M(t_0).
\end{equation*}

If $t\in[t_0,t_0+\tau]$, then we have for any $i_0\in\mathcal{I}_0$,
\begin{equation}\label{eq:the boundary of x_i_0}
x_{i_0}(t)\le \gamma_0x_m(t_0)+(1-\gamma_0)x_M(t_0),
\end{equation}
where $\gamma_0=e^{-\tau\bar{a}}$.

Choose $i_1\in\mathcal{I}_1:=\{i:\exists\,j\in\mathcal{I}_0$, $j\in\mathcal{N}_i\}$. By the conditions of Theorem \ref{th:distance=0} and the equation \eqref{eq:the boundary of x_i_0}, it is trivial to get that $f_{i_0i_1}\big(x_{i_0}(t)\big)<x_M(t_0)$. Hence, for any $t\in[t_0,t_0+\tau/n]$, there exists a constant $\gamma_0'\in(0,1)$ such that
\begin{equation*}
f_{i_0i_1}\big(x_{i_0}(t)\big)\le\gamma_0'x_m(t_0)+(1-\gamma_0')x_M(t_0).
\end{equation*}
Then, we can get that
\begin{align*}
&x_{i_1}(t_0+\frac{\tau}{n})\\
\le& e^{-\alpha_{i_1}\frac{\tau}{n}}x_{i_1}(t_0)+\big[a_{i_1i_0}\gamma_0'\big(x_m(t_0)-x_M(t_0)\big)\\
&+\alpha_{i_1}x_M(t_0)\big]\int_{t_0}^{t_0+\frac{\tau}{n}}e^{-\alpha_{i_1}(t_0+\frac{\tau}{n}-s)}ds\\
\le& e^{-\alpha_{i_1}\frac{\tau}{n}}x_M(t_0)+(1-e^{-\alpha_{i_1}\frac{\tau}{n}})x_M(t_0)\\
&+a_{i_1i_0}\gamma_0'\big(x_m(t_0)-x_M(t_0)\big)\int_{t_0}^{t_0+\frac{\tau}{n}}e^{-\alpha_{i_1}(t_0+\frac{\tau}{n}-s)}ds\\
=& x_M(t_0)+\frac{a_{i_1i_0}}{\alpha_{i_1}}(1-e^{-\alpha_{i_1}\frac{\tau}{n}})\gamma_0'\big(x_m(t_0)-x_M(t_0)\big).
\end{align*}

Since there exists a constant $\rho_1>0$ such that for any $i_1\in\mathcal{I}_1$, $i_0\in\mathcal{N}_{i_1}$, $\rho_1\le \frac{a_{i_1i_0}}{\alpha_{i_1}}(1-e^{-\alpha_{i_1}\frac{\tau}{n}})$. Therefore, it means $x_{i_1}(t_0+\frac{\tau}{n})\le\rho_1\gamma_0'x_m(t_0)+(1-\rho_1\gamma_0')x_M(t_0)$.
Similar to \eqref{eq:the boundary of x_i_0}, we can get that for any $i_1\in\mathcal{I}_1$, $t\in[t_0+\frac{\tau}{n},t_0+\tau]$,
\begin{equation}\label{eq:the boundary of x_i_1}
x_{i_1}(t)\le \gamma_1 x_m(t_0)+(1-\gamma_1)x_M(t_0),
\end{equation}
where $\gamma_1=\rho_1\gamma_0'\gamma_0$. Continuing the above analysis over $[t_0+\frac{\tau}{n}m,t_0+\tau]$, $\forall\,m=1,2,\dots,n-1$, it can be concluded that for all $i\in\mathbf{N}$,
\begin{equation*}
x_i(t_0+\tau)\le\gamma_{n-1}x_m(t_0)+(1-\gamma_{n-1})x_M(t_0),
\end{equation*}
where $\gamma_{n-1}=\rho_{n-1}\gamma_{n-2}'\gamma_{0}$.

If $x_m(t_0)<x_M(t_0)$, then there exists a constant $\omega\in(0,1]$ such that $x_m(t_0)\le\omega\partial+(1-\omega)x_M(t_0)$. Here, we have
\begin{equation*}
Y_1(t_0+\tau)=(1-k_2)\big(x_M(t_0+\tau)-\partial\big)\le(1-\omega\gamma_{n-1})Y_1(t_0).
\end{equation*}

If $x_m(t_0)=x_M(t_0)>\partial_M$, we use the Assumption \ref{assumption_4} to get the convergence. Since there exists $i_0$, $i_1$ such that $f_{i_0i_1}\big(x_{i_0}(t_0)\big)$ is continuous on $x_{i_0}(t_0)=x_M(t_0)$ and $f_{i_0i_1}\big(x_{i_0}(t_0)\big)<x_M(t_0)$. Hence, there exists $T(\omega')$ such that $\forall\,t\in[t_0,t_0+T(\omega')]$, $f_{i_0i_1}\big(x_{i_0}(t)\big)\le \omega'\partial+(1-\omega')x_M(t_0)$, where the constant $\omega'\in(0,1]$. Similar to \eqref{eq:the boundary of x_i_1}, we have $\forall\,t\in[t_0+T(\omega'),t_0+nT(\omega')]$,
\begin{equation*}
x_{i_1}(t)\le \omega_1 \partial+(1-\omega_1)x_M(t_0),
\end{equation*}
where $\omega_1=\rho_1\omega'\gamma_0$. Furthermore, it shows that
\begin{equation*}
Y_1\big(t_0+nT(\omega')\big)\le(1-\omega_{n-1})Y_1(t_0),
\end{equation*}
where $\omega_{n-1}=\rho_{n-1}\omega_{n-2}'\gamma_{0}$.

The analysis of Case \ref{case:Y_1} is completed. The case $Y(t_0)=Y_3(t_0)=L_1\big(x_m(t_0)\big)-x_m(t_0)$ is symmetric to Case \ref{case:Y_1}, so we omit its analysis.
\end{case}

\begin{case}\label{case:Y_2}
$Y(t_0)=Y_2(t_0)=x_M(t_0)-\partial_m$.

By Lemma \ref{le:the boundary of x_M and x_m}, we can get that $\forall\,i\in\mathbf{N}$, $t\ge t_0$, $x_M(t_0)\ge L_1\big(x_i(t)\big)$. Hence, it is trivial to get that
\begin{equation*}
x_i(t_0+\tau)\le\gamma_{n-1}x_m(t_0)+(1-\gamma_{n-1})x_M(t_0),\enspace \forall\,i\in\mathbf{N}.
\end{equation*}
Furthermore, we can get that $Y_2(t_0+\tau)\le(1-\omega\gamma_{n-1})Y_2(t_0)$ or $Y_2\big(t_0+nT(\omega')\big)\le(1-\omega_{n-1})Y_2(t_0)$. The case $Y(t_0)=Y_4(t_0)=\partial_M-x_m(t_0)$ is symmetric to this case, so we omit its analysis.

Finally, we get $x(t)\to[\partial_m,\partial_M]^n$ as $t\to\infty$, and the proof of the Possibility \ref{item:th_distance=0_P1} is completed.
\end{case}
\end{step1}

\begin{step1}\label{th_distance=0_partB_step1}
In this step, we will complete the proof of Possibility \ref{item:th_distance=0_P2}, i.e., $\partial_M-\partial_m<(1-k_2)(\partial_M-\partial)$ or $\partial_M-\partial_m<(1-k_1)(\partial-\partial_m)$. By symmetry, let $\partial_M-\partial_m<(1-k_1)(\partial-\partial_m)$, which implies that $\exists\,\partial'_M>\partial_M$, such that $\partial'_M-\partial_m=(1-k_1)(\partial-\partial_m)$. Denote
\begin{align*}
Y'(t)=&\max\big\{\partial_M'-\partial_m,x_M(t)-\partial_m,\partial_M'-x_m(t),\\
&(1-k_2)\big(x_M(t)-\partial\big),(1-k_1)\big(\partial-x_m(t)\big)\big\}.
\end{align*}

Similar to the proof in Step \ref{th_distance=0_partA_step1}, it concludes that $\mathbf{x}(t)\to[\partial_m,\partial_M']^n$ as $t\to\infty$.
Then, we use a contradiction argument to prove that for any solution $\mathbf{x}(t)\to[\partial_m,\partial_M]^n$ as $t\to\infty$.

Assume that there exist a solution $\hat{x}(t)$ and $i^*\in\mathbf{N}$, such that $\hat{x}_{i^*}(t)\to(\partial_M,\partial_M']$ as $t\to\infty$. Then, there is a $T^*$ such that for any $t>T^*$ and $j\in\mathbf{N}$, $f_{i^*j}\big(\hat{x}_{i^*}(t)\big)>\partial_m$.

Since the directed graph $\mathcal{G}$ is strongly connected, we can get that there exists a $T'>T^*$ such that $\forall\,t>T'$, $i\in\mathbf{N}$, $\hat{x}_i(t)>\partial_m$.

Denote $Z(t)=\max\{\hat{x}_M(t),\partial_M\}$. Repeat the analysis of Step \ref{th_distance=0_partA_step1}, it shows that $\lim\limits_{t\to\infty}Z(t)=\partial_M$, which implies that
\begin{equation*}
\limsup\limits_{t\to\infty}\hat{x}_{i^*}(t)\le\partial_M.
\end{equation*}
Therefore, the trajectory of $\hat{x}_{i^*}(t)$ cannot converge to $(\partial_M,\partial_M']$ as $t\to\infty$. 
Here we prove that for any solution $\mathbf{x}(t)\to[\partial_m,\partial_M]^n$ as $t\to\infty$.

\end{step1}

The proof is completed.
\end{proof}

\begin{remark}
In the sufficiency proof, the idea of constructing auxiliary variables to analyze the boundedness of MAS is inspired by \cite{fontan2019interval}. However, since our dynamics is not Lipschitz continuous, the LaSalle invariance principle is not applicable. We design some linear boundaries and propose the corresponding lemmas to eliminate nonlinearity, and analyze the states' trending to obtain the boundedness of agents' states.
\end{remark}

\section{Proof of Theorem \ref{th:consensus}}
\subsection{Technical lemma}

\begin{Lemma}\label{le:rc}
\emph{\bf{(Proposition 4.10 in \cite{shi2013robust})}}
Let graph $\mathcal{G}$ has a directed spanning tree, and consider the dynamics of MAS defined over $\mathcal{G}$:
\begin{equation*}
\dot{x}_i(t)=\sum^N_{j=1}a_{ij}\big(x_j(t)-x_i(t)\big)+\theta_i(t),\quad i=\mathbf{N},
\end{equation*}
in which $\theta_i(t)$ is piecewise continuous on $[t_0,\infty)$ and is finite. If $\lim\limits_{t\to\infty}\theta_i(t)=0$, $\forall\,i\in\mathbf{N}$, then $\lim\limits_{t\to\infty}x_i(t)-x_j(t)=0$, $\forall\,i,j\in\mathbf{N}$.
\end{Lemma}

\subsection{Proof of Theorem \ref{th:consensus}}
\begin{proof}
\subsubsection{Necessity}
Antagonistic interaction means that the underlying edges between agents have negative weights.
A signed graph $\mathcal{G}_A$ is \textbf{structurally balanced} if there exists a bipartition $\{\mathcal{V}_1,\mathcal{V}_2\}$ of the nodes, where $\mathcal{V}_1\cup \mathcal{V}_2=\mathcal{V}$, $\mathcal{V}_1\cap \mathcal{V}_2=\emptyset$, such that $a_{ij}\ge0\enspace\forall\, v_i,v_j\in\mathcal{V}_m$ $(m\in\{1,2\})$ and $a_{ij}\le0\enspace\forall\, v_i\in\mathcal{V}_m$, $v_j\in\mathcal{V}_l$, $m\ne l$ $(m,l\in\{1,2\})$ \cite{altafini2012consensus}.
We use a contradiction argument.
Consider the MAS with antagonistic interactions
\begin{equation}\label{MAS_3}
\dot{x}_i(t)=\sum^N_{j=1}|a^s_{ij}|\big(\text{sign}(a^s_{ij})\cdot x_j(t)-x_i(t)\big),
\end{equation}
in which the signed graph $\mathcal{G}_s=\{\mathcal{V}_s,\mathcal{E}_s,\mathcal{A}_s=[a^s_{ij}]\}$,
and suppose the signed graph $\mathcal{G}_s$ is strongly connected and structurally balanced. By the bipartite consensus theorem (Theorem 2 in \cite{altafini2012consensus}), it shows that the system \eqref{MAS_3} reaches bipartite consensus but not consensus, in which agents' state values are the same except for the sign.

Let $\mathcal{A}'=|\mathcal{A}_s|=[|a^s_{ij}|]$ and $\mathcal{G}'=\{\mathcal{V}_s,\mathcal{E}_s,\mathcal{A}'\}$, i.e., $\mathcal{G}'$ is a strongly connected graph with only cooperative interactions.
Assume the MAS \eqref{MAS_1} is under the graph $\mathcal{G}'$, and for all $i\in\mathbf{N},\,j\in\mathcal{N}_i$, $f_{ji}\big(x_j(t)\big)=\text{sgn}(a^s_{ij})\cdot x_j(t)$, which implies that $k_1k_2>1$. With the above assumption, the dynamics of MAS \eqref{MAS_1} is equivalent to the dynamics of MAS \eqref{MAS_3}. Therefore, it is turns out that MAS \eqref{MAS_1} cannot achieves consensus.

On the other hand, it is obvious that under the above assumption, the system \eqref{MAS_1} satisfies all conditions in Theorem \ref{th:consensus}. By Theorem \ref{th:consensus}, the states of agents will converge to a consensus value. Hence, we get a contradiction and the proof for the necessity statement of Theorem \ref{th:consensus} is proved.


\subsubsection{Sufficiency}
Applying Theorem \ref{th:distance=0}, we have:
\begin{equation}\label{eq:th_consensus_tending_x}
\mathbf{x}(t)\to[\partial_m,\partial_M]^n,\enspace \text{as }t\to\infty.
\end{equation}

Notice that if $\partial_m=\partial_M$, then it turns out that $\lim\limits_{t\to+\infty}x_i(t)=\partial_m=\partial_M=v^*$, $\forall\, i\in\mathbf{N}$,
and therefore the sufficiency statement of Theorem \ref{th:consensus} is proved.

Hence, we continue our proof in the condition that $\partial_m<\partial_M$. Assume $\partial_m<\partial_M$ in the following.

\begin{step2}\label{step:th_consensus_error=0}
In this step, it shows that the states of agents tend to achieve consensus.

Denote $\theta_i(t)=\sum\limits_{j\in\mathcal{N}_i}a_{ij}\Big(f_{ji}\big(x_j(t)\big)-x_j(t)\Big)$, and the dynamics of MAS \eqref{MAS_1} can be rewritten as
\begin{equation}\label{eq:th_consensus_robust_system}
\frac{d}{dt}x_i(t)=\sum\limits_{j\in\mathcal{N}_i}a_{ij}\big(x_j(t)-x_i(t)\big)+\theta_i(t).
\end{equation}

Then, we use a contradiction argument to prove that $\lim\limits_{t\to\infty}\theta_i(t)=0$, $\forall\,i\in\mathbf{N}$.
Without loss of generality, assume that there exists a solution $x(t)$, such that $\forall\,i\in\mathbf{N}$, $\liminf\limits_{t\to\infty}\theta_i(t)\ge0$ and $\exists\,i^*\in\mathbf{N}$, $\liminf\limits_{t\to\infty}\theta_{i^*}(t)>0$.

$\mathcal{P}=\{\chi_1,\chi_2,\dots\}$ denotes the set of all limit points of $x_{i^*}(t)$ as $t\to\infty$, i.e., there are time sequences $\{t_n'\}$ with $\lim\limits_{n\to\infty}t_n'=\infty$ and $\lim\limits_{n\to\infty}x_{i^*}(t_n')=\chi_1$, and $\{t_n''\}$ with $\lim\limits_{n\to\infty}t_n''=\infty$ and $\lim\limits_{n\to\infty}x_{i^*}(t_n'')=\chi_2$, etc.
Since $\liminf\limits_{t\to\infty}\theta_{i^*}(t)>0$, it turns out that $\exists\,i':i^*\in\mathcal{N}_{i'}$, such that
\begin{equation*}
\sup\limits_{\chi_i\in\mathcal{P}}\Big\{\limsup\limits_{x\to\chi_i}\big(f_{i^*i'}(x)-x\big)\Big\}>0.
\end{equation*}
Furthermore, since $x_{i^*}(t)\to[\partial_m,\partial_M]$ as $t\to\infty$, it can be concluded that $\limsup\limits_{x\to\partial_m}f_{i^*i'}(x)>\partial_m$.

Therefore, there is a time sequence $\{t_n^*\}\to\infty$ with $\{x_{i^*}(t_n^*)\}\to\partial_m$ and $\lim\limits_{n\to\infty}f_{i^*i'}\big(x_{i^*}(t_n^*)\big)>\partial_m$. It is clear that $\partial_m\in\mathcal{P}$.

Since $f_{i^*i'}(x)$ is continuous in $[\partial_m,\partial_M]$ and has finite breaks in $(-\infty,\partial_m)$, there exists a time sequence $\{t_n^\sim\}\subseteq\{t_n^*\}$ and $f_{i^*i'}(x)$ is continuous on $x_{i^*}(t)$, $\forall\,t\in\{t_n^\sim\}$, which implies that $f_{i^*i'}\big(x_{i^*}(t)\big)$ is continuous on $\{t_n^\sim\}$.

Further, there exist $\epsilon>0$ and $T(\epsilon)$, such that $f_{i^*i'}\big(x_{i^*}(t)\big)>\partial_m$ and $f_{i^*i'}\big(x_{i^*}(t)\big)$ is continuous in $(t-\epsilon,t+\epsilon)$, for all $t>T(\epsilon)$ and $t\in\{t_n^\sim\}$.
Denote closed and connected interval $\mathbf{I}_k=[t_k-\frac{\epsilon}{2},t_k+\frac{\epsilon}{2}]$, where $t_k>T(\epsilon)$ and $t_k\in\{t_n^\sim\}$, $k\in\mathbb{N}^+$.

Since $\liminf\limits_{t\to\infty}\theta_i(t)\ge0$, $\forall\,i\in\mathbf{N}$, it can be concluded that $\liminf\limits_{t\to\infty}f_{ij}\big(x_i(t)\big)\ge\partial_m$, $\forall\,i,j\in\mathbf{N}$.

Notice that $\int_{\mathbf{I}_k}f_{i^*i'}\big(x_{i^*}(t)\big)>\partial_m$ and repeat the analysis of Theorem \ref{th:distance=0}, we can get that $\liminf\limits_{t\to\infty}x_{i^*}(t)>\partial_m$.
Then, it turns out that $\partial_m\notin\mathcal{P}$. Here, we find a contradiction.

Hence, we have proven that $\lim\limits_{t\to\infty}\theta_i(t)=0$, $\forall\,i\in\mathbf{N}$.

Applying Lemma \ref{le:rc}, we can get that
\begin{equation}\label{eq:th_consensus_error=0}
\lim\limits_{t\to+\infty}x_i(t)-x_j(t)=0,\quad\forall\, i,j\in\mathbf{N}.
\end{equation}
\end{step2}

\begin{step2}
In Step \ref{step:th_consensus_error=0}, it shows that the states of agents will converge to consensus. In this step, by the fact that $x(t)\to[\partial_m,\partial_M]^n$, we prove that for any $i\in\mathbf{N}$, $\lim\limits_{t\to\infty}x_i(t)=v^*$ and $v^*\in[\partial_m,\partial_M]$.

By \eqref{eq:th_consensus_tending_x}, it turns out that for any $\omega_1>0$, there exists a finite $T_1>0$, which holds the following inequation:
\begin{equation}\label{eq:omega_1}
\partial_m-\omega_1\le x_i(t)\le \partial_M+\omega_1,\quad\forall\, t\ge T_1,\,i\in\mathbf{N}.
\end{equation}

Without loss of generality, assume that $\frac{\partial_m+\partial_M}{2}\le x_k(T_1)\le \partial_M+\omega_1$, where $k$ is a fixed node.

Similarly, by \eqref{eq:th_consensus_error=0}, $\exists\,T_2>0$ which is finite, there holds
\begin{equation}\label{eq:omega_2}
|x_i(t)-x_k(t)|\le\omega_2,\quad\forall\,t\ge T_2,\,i\in\mathbf{N}.
\end{equation}


According to \eqref{eq:omega_1} and \eqref{eq:omega_2}, let $\omega_1$ and $\omega_2$ be sufficiently small, and we get that $\partial_m<x_i(T_*)<\partial_M+\omega_1+\omega_2$, $\forall\,i\in\mathbf{N}$, where $T_*>\max\{T_1,T_2\}$.

Depending on whether $\exists\,l\in\mathbf{N}$, $x_l(T_*)\ge \partial_M$ or not, there are two cases in the following proof.
\begin{enumerate}
\item $\exists\,l\in\mathbf{N}$, $x_l(T_*)\ge \partial_M$.
Repeating the analysis in Step \ref{th_distance=0_partA_step1} in the proof of Theorem \ref{th:distance=0}, we can get that
\begin{equation*}
\bar{Y}(t)=\max\limits_{i\in\mathbf{N}}\Big\{(1-k_2)\big(x_i(t)-\partial\big)\Big\}
\end{equation*}
is non-increasing for $t\ge T_*$. Since $1-k_2>0$, it turns out that $\max\limits_{i\in\mathbf{N}}x_i(t)$ is non-increasing for $t\ge T_*$.

\item $\forall\,i\in\mathbf{N}$, $\partial_m<x_i(T_*)< \partial_M$.
It is easy to get $f_{ij}\big(x_i(T_*)\big)=x_i(T_*)$, $\forall\,i,j\in\mathbf{N}$, and the system degenerates into a standard multiagent system at time $T_*$. Therefore, it is easy to know that $\max\limits_{i\in\mathbf{N}}x_i(t)$ is non-increasing for $t\ge T_*$.

\end{enumerate}

Combining the above analyses, we can conclude that $\max\limits_{i\in\mathbf{N}}x_i(t)$ is non-increasing for $t\ge T_*$. Furthermore, $\max\limits_{i\in\mathbf{N}}x_i(t)$ converges to a finite limit value (denote the value by $\overline{v}$). According to \eqref{eq:th_consensus_error=0}, $\min\limits_{i\in\mathbf{N}}x_i(t)$ must converge to the same limit value $\overline{v}$. Since $\min\limits_{i\in\mathbf{N}}x_i(t)\le x_j(t)\le \max\limits_{i\in\mathbf{N}}x_i(t)$, $\forall\, j\in\mathbf{N}$, it is trivial to get that $\lim\limits_{t\to\infty}x_i(t)=\bar{v}=v^*$, $\forall\, i\in\mathbf{N}$.

Using \eqref{eq:th_consensus_tending_x}, we can conclude that $v^*\in[\partial_m,\partial_M]$.
Based on the above analysis, it is shown that all $x_i(t)$ will converge to a finite limit $v^*$ and $v^*\in[\partial_m,\partial_M]$.
\end{step2}
\end{proof}

\begin{remark}
The robust consensus idea is inspired by \cite{fontan2019interval}. However, since our dynamics is not Lipschitz continuous, the system may have multiple solutions. We analyze the limit points of multiple solutions and integrate the relevant variables over a short period to analyze the system's convergence.
\end{remark}

\section{Proof of Theorem \ref{th:exist_equilibrium}}
\begin{proof}
By Lemma \ref{le:positive_invariant}, it shows $[ \partial_m, \partial_M]^n$ is a positively invariant set.
By the Brouwer fixed point Theorem extended to dynamical systems \cite{basener2006brouwer}, we can conclude that there exists an equilibrium in $[ \partial_m, \partial_M]^n$.
Hence, along the system \eqref{MAS_1}, the existence of equilibria is proven.
By Theorem \ref{th:distance=0}, it shows that
\begin{equation*}
\lim\limits_{t\to\infty}\textnormal{distance}\big([ \partial_m, \partial_M]^n,\mathbf{x}(t)\big)=0,
\end{equation*}
which implies that every equilibrium $\mathbf{e}\in[\partial_m,\partial_M]^n$.
\end{proof}

\section{Proof of Theorem \ref{th:converge_to_equilibrium}}
\begin{proof}
Denote
\begin{align*}
V(t)=&\max\limits_{i\in\mathbf{N}}\big\{(1-k_{e1})\big(\mathbf{e}_i-x_i(t)\big),(1-k_{e2})\big(x_i(t)-\mathbf{e}_i\big)\big\}\\
=&\max\limits_{i\in\mathbf{N}}\big\{(1-k_{e1})\big(-\varepsilon_i(t)\big),(1-k_{e2})\varepsilon_i(t)\big\},
\end{align*}
and clearly $V$ is Lipschitz continuous. At first, we will prove that $V(t)$ is a non-increasing function.

Denote $\varepsilon_m(t)=\min\limits_{i\in\mathbf{N}}\varepsilon_i(t)$, $\varepsilon_M(t)=\max\limits_{i\in\mathbf{N}}\varepsilon_i(t)$.

By the structure of $V(t)$, there exists two cases:
\begin{enumerate}
\item\label{it:th_converge_to_equilibrium_case1} $V(t)=(1-k_{e2})\varepsilon_M(t)$;
\item\label{it:th_converge_to_equilibrium_case2} $V(t)=(1-k_{e1})\big(-\varepsilon_m(t)\big)$.
\end{enumerate}

We first consider the Case \ref{it:th_converge_to_equilibrium_case1}, which implies that $\exists\,t^*$, $V(t^*)=\max\limits_{i\in\mathbf{N}}\big\{(1-k_{e2})\varepsilon_i(t^*)\big\}$ and $\varepsilon_M(t^*)>0$.
Denote $\mathcal{I}_e(t)=\big\{k:\varepsilon_k(t)=\max\limits_{i\in\mathbf{N}}\varepsilon_i(t)\big\}$. By Lemma \ref{le:der_to_max}, we have
\begin{align*}
&d^+V(t^*)=d^+\max\limits_{i\in\mathbf{N}}\Big\{\big(1-k_{e2}\big)\varepsilon_i(t^*)\Big\}\notag\\
=&\max\limits_{i\in\mathcal{I}_e(t^*)}\Big\{(1-k_{e2})\sum\limits_{j\in\mathcal{N}_{i}}a_{ij}\Big(f_{ji}\big(x_j(t^*)\big)-x_{i}(t^*)\Big)\Big\}\notag\\
=&\max\limits_{i\in\mathcal{I}_e(t^*)}\Big\{(1-k_{e2})\sum\limits_{j\in\mathcal{N}_{i}}a_{ij}\Big(f_{ji}\big(\mathbf{e}_j+\varepsilon_j(t^*)\big)-\mathbf{e}_{i}-\varepsilon_{i}(t^*)\Big)\Big\}.
\end{align*}
Furthermore, noticing that $\sum\limits_{j\in\mathcal{N}_{i}}a_{ij}\big(f_{ji}(\mathbf{e}_j)-\mathbf{e}_{i}\big)=0$, we can conclude that
\begin{align}\label{eq:asymptotically stabel th_case_i}
\dot{\varepsilon_{i}}(t^*)=\sum\limits_{j\in\mathcal{N}_{i}}a_{ij}\big(f_{ji}\big(\mathbf{e}_j+\varepsilon_j(t^*)\big)-f_{ji}(\mathbf{e}_j)-\varepsilon_{i}(t^*)\big).
\end{align}

Let $\varepsilon_{i'}=\varepsilon_M$, which implies $i'\in\mathcal{I}_e(t^*)$.
Applying Lemma \ref{le:x_M and x_m} on \eqref{eq:asymptotically stabel th_case_i} and under condition 2, we can get that
\begin{equation*}
\begin{cases}
\varepsilon_{i'} \ge \varepsilon_j\ge f_{ji'}\big(\mathbf{e}_j+\varepsilon_j\big)-f_{ji'}(\mathbf{e}_j) &\text{if } \varepsilon_j\ge0;\\
\varepsilon_{i'} \ge L_{e1}(\varepsilon_j)> f_{ji'}\big(\mathbf{e}_j+\varepsilon_j\big)-f_{ji'}(\mathbf{e}_j) &\text{if } \varepsilon_j<0.
\end{cases}
\end{equation*}

By the above two analysis, it concludes that $d^+V(t^*)\le0$ when $(1-k_{e2})\varepsilon_M(t^*)>(1-k_{e1})\big(-\varepsilon_m(t^*)\big)$.

The proof of Case \ref{it:th_converge_to_equilibrium_case2} is similar to the above proof, and hence are omitted here.
Combining the two cases, we have proved that $d^+V(t)\le0$ for all $t\ge t_0$.

Repeating the analysis of Theorem \ref{th:distance=0}, we have $\mathbf{x}(t)\to\mathbf{e}$, as $t\to\infty$.
Since the equilibrium $\mathbf{e}$ is asymptotically stable and the initial states $\mathbf{x}(t_0)$ can be arbitrary, $\mathbf{e}$ is a unique equilibrium.
The proof is completed.
\end{proof}

\section{Proofs of Theorem \ref{th:single equilibrium} and Corollary \ref{co:single equilibrium}}
\subsection{Proof of Theorem \ref{th:single equilibrium}}
\begin{proof}
We first prove the existence of equilibria.

By conditions 2 and 3, we can get that for any $j\in\mathbf{N},\,i\in\mathcal{N}_j$ and $\omega\ne0$,
\begin{equation}\label{eq:f_ij is continuous}
-1<k^*\le\frac{f_{ij}(x+\omega)-f_{ij}(x)}{\omega}\le1,\quad x\in\mathbb{R},
\end{equation}
which implies that $f_{ij}$ is a continuous function. For simplicity, let $k^*\in(-1,0)$.

By \eqref{eq:f_ij is continuous}, it can be concluded that there exists at least one intersection between functions $f_{ij}$ and $f(x)=x$, i.e., $\Theta_{ij}\ne\emptyset$. Denote $X_M=\max\big\{x:x\in\bigcup\Theta_{ij}\big\}$ and $X_m=\min\big\{x:x\in\bigcup\Theta_{ij}\big\}$.
Since $\Theta_{ij}\ne\emptyset$, let $x_{ij}^*\in\Theta_{ij}$, i.e., $f_{ij}(x_{ij}^*)=x_{ij}^*$. By \eqref{eq:f_ij is continuous}, it shows that
\begin{equation*}
\left\{
\begin{aligned}
f_{ij}(x_{ij}^*+\omega)\le f_{ij}(x_{ij}^*)+\omega=x_{ij}^*+\omega,\quad \omega>0;\\
f_{ij}(x_{ij}^*+\omega)\ge f_{ij}(x_{ij}^*)+\omega=x_{ij}^*+\omega,\quad \omega<0,
\end{aligned}
\right.
\end{equation*}
which implies that
\begin{equation}\label{eq:f_ij range1}
\left\{
\begin{aligned}
f_{ij}(x)\le x,\quad x\ge X_M;\\
f_{ij}(x)\ge x,\quad x\le X_m.
\end{aligned}
\right.
\end{equation}

There are two parallel lines with slope $k^*\in(-1,0)$:
\begin{equation*}
\left\{
\begin{aligned}
L_m(x)&=k^*x+(1-k^*)X_m,\\
L_M(x)&=k^*x+(1-k^*)X_M,
\end{aligned}
\right.
\end{equation*}
and it can be concluded that
\begin{equation*}
\left\{
\begin{aligned}
f_{ij}(x)\le L_M(x),\quad x\le X_M;\\
f_{ij}(x)\ge L_m(x),\quad x\ge X_m.
\end{aligned}
\right.
\end{equation*}

Let $L^*(x)=-x+X_M+X_m$ with slope $k=-1$.

Since $-1<k^*<0$, it is easy to know that between $L^*$ and the parallel lines $L_M,L_m$ exist two intersections $(y_m,y_M)$ and $(y_M,y_m)$ where $y_M\ge y_m$, which implies that $L_M(y_m)=y_M$ and $L_m(y_M)=y_m$. And it is easy to know that $y_m<X_m\le X_M<y_M$.
Then, it can be concluded that
\begin{equation}\label{eq:f_ij range2}
\left\{
\begin{aligned}
f_{ij}(x)\le L_M(x)\le y_M,\quad y_m\le x\le X_M;\\
f_{ij}(x)\ge L_m(x)\ge y_m,\quad X_m\le x\le y_M.
\end{aligned}
\right.
\end{equation}

Combine \eqref{eq:f_ij range1} and \eqref{eq:f_ij range2}, we can get that $y_m\le f_{ij}(x)\le y_M$, $y_m\le x\le y_M$.
By Theorem \ref{th:exist_equilibrium}, it shows that the system \eqref{MAS_1} has at least one equilibrium.
Then, we will show that the MAS \eqref{MAS_1} has only one asymptotically stable equilibrium.

Assume one of equilibria is $\mathbf{e^*}=\{\mathbf{e}_1^*,\dots,\mathbf{e}_n^*\}^T$, and denote the error between $x(t)$ and $\mathbf{e^*}$ by $\varepsilon_i(t)^*=x_i(t)-\mathbf{e}_i^*$. By \eqref{eq:f_ij is continuous}, we have for all $j\in\mathbf{N},\,i\in\mathcal{N}_j$,
\begin{align*}
\varepsilon_i^*\le f_{ij}(\mathbf{e}_i+\varepsilon_i^*)-f_{ij}(\mathbf{e}_i)\le k^*\varepsilon_i^*,\quad\varepsilon_i^*<0;\\
k^*\varepsilon_i^*\le f_{ij}(\mathbf{e}_i+\varepsilon_i^*)-f_{ij}(\mathbf{e}_i)\le \varepsilon_i^*,\quad\varepsilon_i^*>0.
\end{align*}
Since $k^*\in(-1,0)$, it shows that $k^*k^*<1$. Hence, the Condition 3 of Theorem \ref{th:converge_to_equilibrium} holds.
Because $\bigcap\limits_{(v_j,v_i)\in \mathcal{E}}\Theta_{ij}=\emptyset$ and for any $j\in\mathbf{N},\,i\in\mathcal{N}_j$ and $\omega\ne0$,
\begin{equation*}
-1<\frac{f_{ij}(x+\omega)-f_{ij}(x)}{\omega}<1,\quad x\notin\Theta_{ij},
\end{equation*}
we can conclude that for any $\varepsilon^*\ne0$, there exist $j\in\mathbf{N},i\in\mathcal{N}_j$ and $\delta>0$, such that $f_{ij}(\mathbf{e}_i+\varepsilon')-f_{ij}(\mathbf{e}_i)\ne \varepsilon'$, $\forall\,\varepsilon'\in(\varepsilon^*-\delta,\varepsilon^*+\delta)$.
Hence, the Condition 4 of Theorem \ref{th:converge_to_equilibrium} holds.

Apply Theorem \ref{th:converge_to_equilibrium}, it shows that the equilibrium $\mathbf{e}$ is a unique, asymptotically stable equilibrium.
\end{proof}

\subsection{Proof of Corollary \ref{co:single equilibrium}}
\begin{proof}
Because $f_{ij}(x_i)$ is a continuous and piecewise linear function with its slopes $k_{ij}\in(-1,1]$ and $m_{ij}=0$ when $k_{ij}=1$, it turns out that for any $\omega\ne0$, $-1<\frac{f_{ij}(x+\omega)-f_{ij}(x)}{\omega}<1$, $x\notin\Theta_{ij}$.
Apply Theorem \ref{th:single equilibrium}, this corollary is proved.
\end{proof}

%

\bibliographystyle{IEEEtran}
\bibliography{TCC}

\end{document}